%% file: hel_rev.tex
\documentclass[11pt]{article}  
\input{head}

\begin{document} 
\title{The Effect of Helicity on the Effective Diffusivity for 
Incompressible Random Flows} 
\author{\\ 
{\bf D.S. Dean}\\ \\ IRSAMC, Laboratoire de Physique Quantique, \\ 
Universit\'e Paul Sabatier, 118 route de Narbonne, 31062 Toulouse 
Cedex \\ \\ \\ 
{\bf I.T. Drummond and R.R. Horgan}\\ \\ DAMTP, CMS\\ 
Wilberforce Road, Cambridge CB3 0WA\\} 
\maketitle 
\begin{abstract} 
The advection of a passive scalar by a quenched (frozen) incompressible 
velocity field is studied by extensive high precision numerical 
simulation and various approximation schemes. We show that second 
order self consistent perturbation theory, in the absence of 
helicity, perfectly predicts the effective diffusivity of a tracer 
particle in such a field. In the presence of helicity in the flow 
simulations reveal an unexpectedly strong enhancement of the effective 
diffusivity which is highly nonperturbative and is most 
visible when the bare molecular diffusivity of the particle is 
small. We develop and analyse a series of approximation schemes which 
indicate that this enhancement of the diffusivity is due to a novel 
second order effect whereby the helical component of the field, which 
does not directly renormalize the effective diffusivity, enhances the 
strength of the non helical part of the flow, which in turn 
renormalizes the molecular diffusivity. We show that this 
renormalization is most important at low bare molecular diffusivity, in 
agreement with the numerical simulations. 
 
\end{abstract} 
\vfill DAMTP-2000-139 
\newpage 
\newsection{\label{INTR}\bf Introduction} The advection of passive 
fields subject to molecular diffusion and convection by turbulent 
fluid has been extensively studied by both theoretical and 
computational techniques 
\c{Kra:1,Kra:2,Kra:3,DruDuaHor:1,DruHor:1,DeaDruHor:1}. By comparing 
the results of simulation with the theoretical prediction for various 
long-range quantities, the efficacy of the theoretical methods can be 
tested albeit, in somewhat artificial models. The applications to the 
physics of complex systems and engineering are many fold. In practical 
problems we need to calculate the bulk properties of random media 
given statistical models for the disorder present. In general the 
complexity of these real world problems means that one must resort to 
approximation schemes to calculate these large scale bulk 
properties. It is therefore essential to verify various methods of 
analysis on model problems before one can be confident that these or 
similar methods can be applied to more realistic systems. The success 
of an approach depends on whether the approximation preserves the 
essence of the physical mechanism responsible for determining the 
long-range parameters of the advection in terms of the parameters 
describing the local characteristics of the flow. In this paper
we consider advection in a helical Gaussian turbulent flow  
which was originally studied in \c{DruDuaHor:1}. The surprising result 
observed on the basis of simulation is that the long-range effective 
diffusivity, $\k_e$, is greatly enhanced by the presence of the helicity
by more than a factor of two, the effect being strongest for small molecular 
diffusivity, $\k_0$. In the absence of helicity the calculation of $\k_e$ to 
two loops in self-consistent perturbation theory agrees accurately with 
the simulation for all $\k_0$. However, such an approach predicts that 
even maximal helicity will have only a small effect of the order of 10\%. 
This is in stark contrast to the results of simulation. The puzzle is to explain 
these results for what is a relatively simply posed system. A 
successful theoretical approach will involve infinite resummations of 
contributions and it is in this sense that the enhancement is non-perturbative. 

In this paper we discuss a possible resolution of the conflict between
theory and simulation by using various methods to identify the
low-wavenumber effective theory governing the diffusive dispersal of
particles advected in the turbulent flow when helicity is present.
The derivation of the effective theory is guided by the renormalization
group (RG) idea that the Green function at low wave-number is, 
in some approximation, the solution to an effective second-order
differential equation whose parameters are determined self-consistently
in terms of the original or `bare' defining the model.  The effect of
helicity in the flow causes the turbulent velocity field, $\bu(\bx,t)$, 
to be additively renormalized by a term proportional to the vorticity,
$\bom = \nabla\times\bu$. The coefficient of proportionality is a
is a pseudo-scalar which is generated by the axial-vector nature of 
the helical flow and so depends on the helicity $h$, defined in 
terms of $\bu$ by
\ben
h = \la \bu\cdot\nabla\times\bu \ra~\label{H}
\een
where
$\la \cdot \ra$ denotes the ensemble average over the random
velocity field. In our model the magnitude of $h$ is measured
by a parameter $\l$, $0 \le \l \le 1$, and the results are
given in terms of $\l$. The usual perturbative result for the dependence of 
$\k_e$ on $\l$ is that $\k_e$ is a series in $\l^2$ for all values of 
$\k_0$. This is self-evident since the magnitude of $\k_e$ is 
independent of the sign of $\l$. The simulation is seemingly 
consistent with this fact for $\l < 0.2$ at $\k_0=0$ but is not well
fitted by any simple approach, and for larger $\l$ the curve  
lies far higher than the naive calculation. We discuss an
improved self-consistent scheme which expresses the Green function and 
vertex functions as solutions to integral equations which are solved 
in a low-wavenumber approximation. This method leads to a strong enhancement
of $\k_e$ for increasing $\l$ and, as such, is a good indication that 
we are on the right track. However, for small $\l$ the effect is 
paradoxically too strong, leading to a non-analytic dependence of $\k_e$ on
$\l$ which is predicted to be $\k_e \sim \l^{2/3}$ in the one-loop case. 
This is possibly due to the approximation made in obtaining the solution
but it is a complex matter to ascertain whether this is so. An alternative 
approach is to use the functional Hartree-Fock method which leads to an 
integral equation for the Green function self-energy as a function of 
wave-number. The result of this method for $\k_e(\l)$ is better behaved
at small $\l$ but the predicted enhancement is not big enough and does not
fit the simulation data. In general, the effect is most pronounced for
small $\k_0$ and empirically from our simulation we find that the results 
distinguish the regions $\k_0 \ll 0.2$ and $\k_0 \gg 0.2$. There is a 
pronounced dip in $\k_e$ vs $\k_0$ at $\k_0 \sim 0.2$ for $\l = 1$. This dip is 
not predicted by either of the methods mentioned so far. 

We also present a renormalization group approach which shows a 
mechanism for the enhancing effect of helicity on $\k_e$.  The 
renormalization group is normally most useful for computing anomalous 
exponents since they are generally independent of much of the details 
defining the model: the idea of universality. It is much more 
difficult to control a standard RG analysis if it is used to calculate 
the coefficients of scaling behaviour, i.e., observables like 
$\k_e$. However, in ref. \c{DeaDruHor:2} we reported on a successful 
use of the RG in predicting $\k_e$ for gradient flows and believe that 
an RG analysis can generally give a strong indication of the kind of 
mechanism which influences the size of parameters controlling the 
large-scale characteristics of advection. In this paper we show that 
the flow at large wave vector can strongly enhance $\k_e$ when $\k_0$ 
is small. In particular, this approach does provide a mechanism for
the dip observed in $\k_e$ vs $\k_0$ at $\k_0 \sim 0.2$ for $\l=1$.
 
In section \ref{MODEL} the model and the formalism are reviewed; in 
section \ref{PT} the perturbation theory is briefly described; in 
section \ref{SCM} the self-consistent integral equations for the Green 
function and vertex functions are derived to one-loop and the small 
wave vector approximation for $\k_e$ is derived; in section \ref{HFOCK} 
the functional Hartree Fock method is examined; in section \ref{RG} 
the renormalization group approach is explained and in section 
\ref{CONC} the conclusions are presented. 
 
\newsection{\label{MODEL}The Model and Formalism} In \c{DruDuaHor:1} 
the problem of a passive scalar advected by an incompressible 
turbulent flow with a molecular diffusivity was studied. The turbulent 
fluid velocity field, $\bu(\bx,t)$, was described by its statistical 
properties which were assumed to be Gaussian and so fully determined 
by the velocity auto-correlation function. In the original study  
the flow was time-dependent, but since the enhancement of $\k_e$ 
by helicity in the flow is present also for time-independent flows 
we assume here, for simplicity, a time-independent flow (i.e. quenched or 
frozen turbulence) for which the auto-correlation function can be 
expressed in the following form:
\ben
\la 
u_i(\bx)u_j(\bxp) \ra = \int\dk 
e^{i\bk\cdot(\bx-\bxp)} F_{ij}(\bk)~.\label{ACF} 
\een

The ensemble of velocity fields was taken to be homogeneous and 
isotropic and so for incompressible fluids $F_{ij}(\bk)$ can be 
written as 
\ben
F_{ij}(\bk) = \Phi(\bk)(k^2\de_{ij}-k_ik_j)~+~ 
\Psi(\bk)i\e_{imj}k_m~,\label{SPECTRUM0} 
\een
where $\Psi$ 
represents the presence of helicity in the flow. In \c{DruDuaHor:1} it 
was assumed that $\Phi$ and $\Psi$ took the factorized forms: \bea 
\Phi(\bk)&=&{(2\pi)^3\over 3}A^2 E(k) \nn\\ 
\Psi(\bk)&=&{(2\pi)^3\over 3}A^2 kE(k)\sin 2\psi 
~~,\label{SPECTRUM1} \eea where $A$ is chosen so that 
\ben
\int d\bk 
E(k) = 1~,~~~~~~~\la \bu\cdot\bu \ra = u_0^2~, \label{NORMS} 
\een
and where $u_0$ is the 
r.m.s. velocity. Choosing the angle $\psi$ to be $k$-independent means 
that the helicity is of equal strength at all wave vectors. The 
helicity parameter, $h$, has been defined in eqn. (\ref{H}) and with 
the definitions in eqn. (\ref{SPECTRUM1}), we find 
\ben
h = {2\over 
3}A^2\la k^3 \ra \sin 2\psi~,\label{HPSI} 
\een
where $\la k^3 \ra$ is 
the expectation value with respect to the distribution $E(k)$.  The 
passive scalar field $\Th(\bx,t)$ is advected according to the 
equation 
\ben
{\d\Th \over \d t} = \k_0 \nabla^2\Th - 
\nabla\cdot(\bu\Th)~,\label{DIFFEQ} 
\een
and the effective, or 
long-range, diffusivity, $\k_e$, is defined by \bea \la \bx\cdot\bx 
\ra(t)&=&\la \int d^3x~\bx\cdot\bx~\Th(\bx,t) \ra~,\nn\\ &=&6\k_e t + 
O(t^0)~~\hbox{as}~~t \rightarrow \infty~,\label{KEFF} \eea where $\Th$ 
is normalized to unity: 
\ben
\int d^3x~\Th(\bx,t) = 1~. \label{THNORM} 
\een
 
For the purposes of numerical simulation a particular member of the 
velocity-field ensemble is then realized by 
\c{Kra:1,Kra:2,DruDuaHor:1} \bea 
\bu(\bx)&=&A\sum_{n=1}^N\left\{\bb\bxi_n\:\cos \psi - \bchi_n \wedge 
\bkhat_n\:\sin \psi\eb \wedge \bk_n~\cos (\bk_n\cdot\bx)\right. \nn\\ 
&&~~~~+~\left.\bb\bchi_n\:\cos \psi + \bxi_n \wedge \bkhat_n\:\sin 
\psi\eb \wedge \bk_n~\sin (\bk_n\cdot\bx)~\right\}~, \label{USERIES} 
\eea where the vectors $\xi_n$ and $\chi_n$ are distributed uniformly 
and independently over the unit sphere and the wave vector $\bk_n$ is 
distributed according to the distribution $E(k)$. For $N$ sufficiently 
large the central limit theorem guarantees that $\bu(\bx)$ is Gaussian 
up to $O(1/N)$ corrections. We have used $N=64$ for which these effects 
are sufficiently small for our purposes. 
 
To simulate the evolution of the scalar field $\Th(\bx,t)$ we 
integrate numerically the stochastic equation for the evolution of a 
particle with path $\bx(t)$ given by 
\ben
\dot{\bx}(t) = 
\bu(\bx(t))~+~\bet(t)~, \label{STOCH} 
\een
where $\bet(t)$ is a 
Gaussian random variable with $\la \bet(t) \ra = 0$ and $\la 
\bet(t)\cdot\bet(t^\prime) \ra = 2\k_0\de(t-t^\prime)$. The resulting 
probability distribution for particle position $\bx(t)$ is then 
$\Th(\bx,t)$ with the initial condition $\Th(\bx,0) = \de(\bx)$. 
 
The discrete form of eqn. (\ref{STOCH}) suitable for numerical 
integration is: 
\ben
\bx_{n+1} - \bx_n = \bu(\bx_n)\:\D t~+~(2\k_0\D 
t)^\half\:\beps_n~, \label{DISC_STOCH} 
\een
where $\beps_n$ is a 
Gaussian random three-vector of zero mean and unit variance for each 
component. This equation models eqn. (\ref{STOCH}) correctly to $O(\D 
t)$ but the details of a third-order Runge-Kutta scheme correct to 
$O(\D t^3)$ are given in \c{DruDuaHor:1}. We use this third-order 
scheme in our numerical simulation. 
 
The effective diffusivity, $\k_e$, is then computed from the ensemble 
of paths by \bea \la \bx(t)\cdot\bx(t) \ra_{\hbox{paths}}&=& \lim_{M 
\rightarrow\infty} {1 \over M} \sum_{a=1}^M 
\bx^{(a)}(t)\cdot\bx^{(a)}(t)~,\nn\\ &=& 
6\k_et~+~O(1)~~~~~~~~\hbox{as}~t \rightarrow\infty~. \label{KE_DEF} 
\eea Here $M$ is the total number of paths averaged over and $(a)$ 
label the member of the ensemble of paths. In practice $M$ is finite 
but large enough to give an estimate for $\k_e$ with small error. In 
addition $t$ must be large enough so that the path evolution is in the 
asymptotic regime where the evolution can be suitably described in 
terms of long range effective, or ``renormalized'' quantities. That 
$t$ is large enough is tested by ensuring that the estimate for $\k_e$ 
is independent of $t$ within statistical errors. 
 
\newsection{\label{PT}\bf Perturbation Theory} 
 
The perturbative approach to solving eqn. (\ref{DIFFEQ}) is well known 
\c{PhyCur:1,DruDuaHor:2,DeaDruHor:1} and we only summarize here the 
necessary results. 
 
Since we are interested in the effective parameters governing the 
evolution of the distribution $\Th(\bx,t)$, we study the related Green 
function $G(\bx)$ which satisfies 
\ben
\k_0 \nabla^2 G - 
\bu\cdot\nabla G~=~-\de(\bx)~,\label{GREENEQ} 
\een
where the 
incompressibility of $\bu$ has been used. A perturbation series in 
$\bu/k_0$ for $\Gt(\bk)$ can be generated by iterating the formal 
solution to eqn. (\ref{GREENEQ}) in Fourier space: 
\ben
\Gt(\bk)~=~{1\over \k_0\bk^2}~+~{1\over \k_0\bk^2}\int \dq 
i(\bk-\bq)\cdot\ut(\bq) \Gt(\bk-\bq)~.\label{GMOM} 
\een
The Green 
function averaged over the velocity ensemble, $\la \Gt(\bk) \ra$, can 
then be written as 
\ben
\la \Gt(\bk) \ra~=~{1\over \k_0\bk^2 - 
\S(\bk)}~,\label{GREENAV} 
\een
where the averaging over the velocity 
ensemble is done using Wicks theorem to give a diagrammatic expansion 
and $\S(\bk)$ is given by one particle-irreducible diagrams. The 
asymptotic behaviour in eqn. (\ref{KE_DEF}) implies that the small 
$\bk$ behaviour of $\la \Gt \ra$ is given by 
\ben
\k_e~=~\k_0~-~{\d\over \d\bk^2}\S(\bk)|_{\bk=0}~.  \label{KEFF0} 
\een

\begin{figure}[h] 
\begin{center} 
\leavevmode \epsfxsize=14 truecm\epsfbox{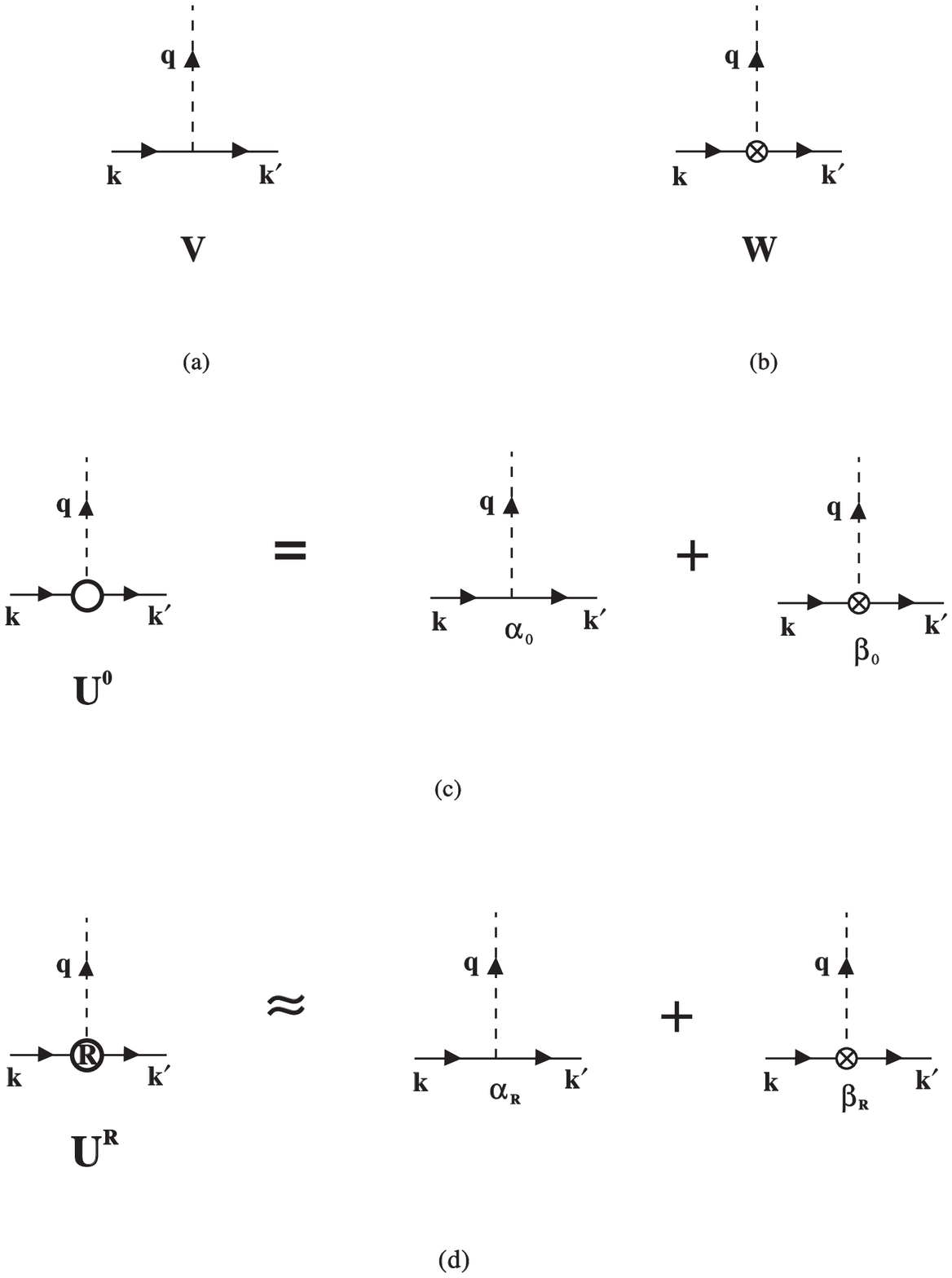} 
\end{center} 
\caption[]{\label{VERTEX}\small The vertices occurring in the 
perturbation schemes: (a) the primitive velocity field vertex; (b) the 
primitive vorticity vertex; (c) the bare complete vertex of the 
effective diffusion equation; (d) the renormalized complete vertex 
approximated as a sum of renormalized vertices associated with the 
velocity field and the vorticity.} 
\end{figure}

The Feynman rules for the diagrammatic perturbation expansion are: 
\bit 
\item[(i)] Wavevector is conserved at each vertex; 
\item[(ii)] Each full line carries a factor of $1/\bk^2~$; 
\item[(iii)] Wavevector is integrated around closed loops with a 
factor $\hbox{d}\bq/(2\pi)^3~$; 
\item[(iv)] The primitive vertex $\bV_i(\bkpr,\bk)$, whose 
diagrammatic representation is shown in figure \ref{VERTEX}a, is given 
by $\bV_i(\bkpr,\bk) = i\:\bkpr_i~$; 
\item[(v)] Each internal dotted line carries a factor 
\ben
F_{ij}(\bk) 
= \Phi(\bk)(k^2\de_{ij}-k_ik_j)~+~ 
\Psi(\bk)i\e_{imj}k_m~.\label{VPROP} 
\een
\eit In what follows we use 
the explicit spectra \bea \Phi(\bk)~&=&~{(2\pi)^{3/2}\over 
6k_0^2}\:u_0^2\,e^{-\bk^2/2k_0^2}~,\nn\\ 
\Psi(\bk)~&=&~\l\:k\:\Phi(\bk)~, \label{SPECTRA} \eea where $\l = \sin 
2\psi~$. 
 
\begin{figure}[h] 
\begin{center} 
\leavevmode \epsfxsize=14 truecm\epsfbox{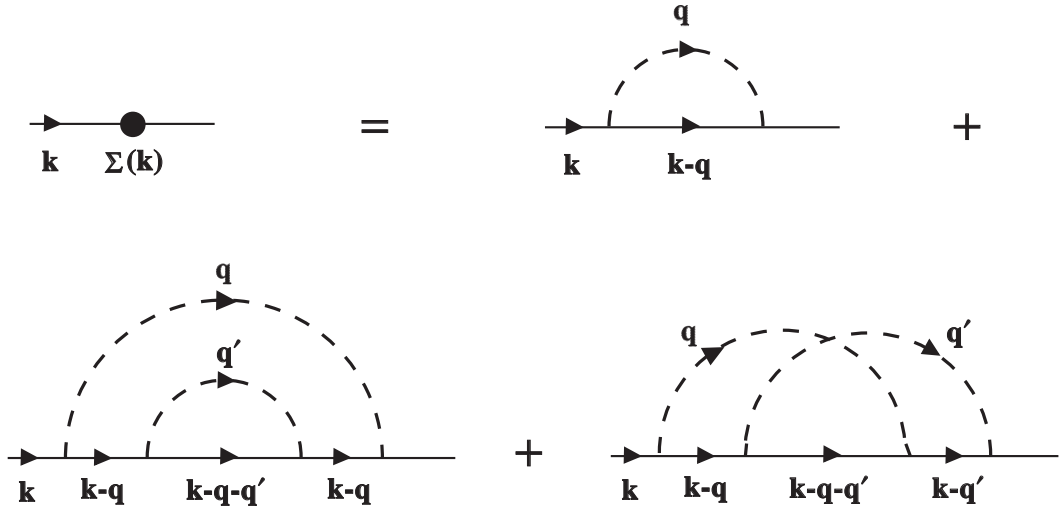} 
\end{center} 
\caption[]{\label{KE_SIMPLE}\small The graphs that contribute to 
two-loop simple perturbation theory for $\Gt(\bk)$.} 
\end{figure}

The simple two-loop calculation for $\k_e$ gives the result 
\ben
\k_e~=~\k_0\bb 1~+~{1\over 9}{u_0^2\over \k_0^2k_0^2}~+~ (0.0059\l^2 - 
0.00884){u_0^4\over \k_0^4k_0^4}\eb~. \label{2LOOP_SIMPLE} 
\een
The 
diagrams contributing to this order are shown in figure 
\ref{KE_SIMPLE}.  The two-loop integrals were done numerically. 
 
The effect of helicity is not seen until second order. This is evident 
from the explicit expressions for the diagrams but is also easily 
understood because the effect of helicity on $\k_e$ cannot depend on 
the sign of $\l$. Hence, the graphs with a non-vanishing contribution 
from helicity must contain an even number of internal velocity 
correlators (dotted lines).

Clearly, this approach is not applicable to the limit $\k_0 
\rightarrow 0$ in which we are interested but a self-consistent method 
will allow the model to be perturbatively analyzed in this limit and 
this is described in the next section. 
 
\newsection{\label{SCM}\bf Self-Consistent Methods} 
 
A self-consistent treatment performs a resummation of an infinite 
subset of diagrams which gives a continuation of the perturbation 
theory beyond its normal radius of convergence. The approach is not 
unique but depends on how the effective low energy theory is 
parametrized and which quantities are treated self-consistently.  A 
successful result will depend on how well the method captures the 
dominant effects in this way. 
 
We first discuss the simplest approach which treats only $\k_e$ 
self-consistently. At two-loops this gives an excellent fit for $\k_e$ 
when $\l=0$ but fails for $\l \ne 0$. We then generalize the method 
and show that we can qualitatively explain the large enhancement in 
$\k_e$ due to helicity although the approach is still quantitatively 
deficient.  Further generalizations are discussed but have not yet 
been carried out.

\newsubsection{\label{SCke} Self-Consistency in $\k_e$} To generate 
the self-consistent perturbation series in $\k_e$ the 
eqn. (\ref{GREENEQ}) for $G(\bx)$ is formally rearranged to become 

\begin{figure}[h] 
\begin{center} 
\leavevmode \epsfxsize=14 truecm\epsfbox{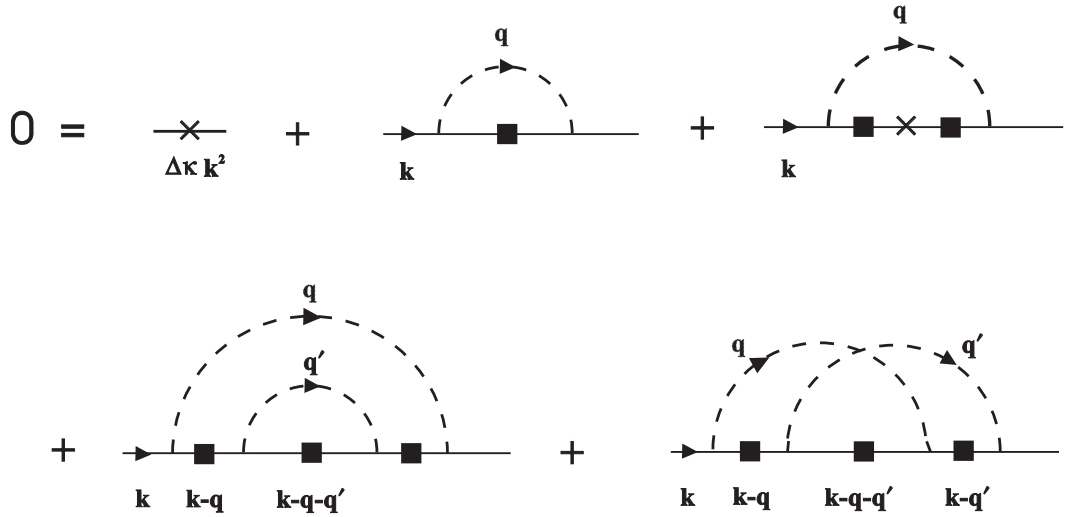} 
\end{center} 
\caption[]{\label{KE_SC}\small The perturbation expansion to two-loops 
of the self-consistent relation for $\k_e$} 
\end{figure}

\ben
\k_e \nabla^2 G - \D \k\: \nabla^2 G - \bu\cdot\nabla 
G~=~-\de(\bx)~,\label{GREENEQ0} 
\een
where $\D \k = \k_e - \k_0~$. The 
second term is a counter-term which is included as part of the 
perturbation. It is formally of first order in the expansion parameter 
which allows the expansion for $\k_e$ to be constructed to a 
consistent order.  The self-consistent perturbation series is 
generated by iterating 
\ben
\Gt(\bk)~=~{1\over \k_e\bk^2}~+~{1\over 
\k_e\bk^2}\int \dq \bb i(\bk-\bq)\cdot\ut(\bq) - \D \k\: \de (\bq)\: 
(\bk-\bq)^2\eb \Gt(\bk-\bq)~.  \label{GMOM_SC} 
\een
This equation can 
be rewritten as an equation for $\D\k$ and its diagrammatic 
representation up to two loops is shown in figure 
\ref{KE_SC}(a). Since $\k_e$ is not renormalized from the tree-level 
value we have the self-consistency condition 
\ben
{\d\over 
\d\bk^2}\S(\bk)|_{\bk=0}~=~0~. \label{SC_COND} 
\een
To $N$-th order in 
$u_0^2/\k_e^2k_0^2$ it is always possible to write this condition in 
the form \bea \k_e&=&\k_0~+~\k_eF_N(\k_e,\l)~, \nn\\ 
F_N(\k_e,\l)&=&\sum_{n=1}^N g_n(\l)\:\bb{u_0^2 \over 
\k_e^2k_0^2}\eb^n~.\label{SC_EQ} \eea 
 
>From now on we set $u_0 = 1,~k_0 = 1$ and, using the velocity-field 
spectrum given in section \ref{MODEL}, we find the two-loop 
self-consistent expression for $\k_e$ 
\ben
\bb\k_e~-~\k_0\eb\bb 
1~-~{1\over 9\:\k_e^2}\eb~=~\k_e\bb {1\over 9\k_e^2}~+~ {1\over 
\k_e^4}\bbs 0.0059\l^2~-~0.00884\ebs \eb~. \label{2LOOP_SC} 
\een
This 
result can be re-expressed in the form of eqn. (\ref{SC_EQ}) to become 

\begin{figure}[h] 
\begin{center} 
\leavevmode \epsfxsize=14 truecm\epsfbox{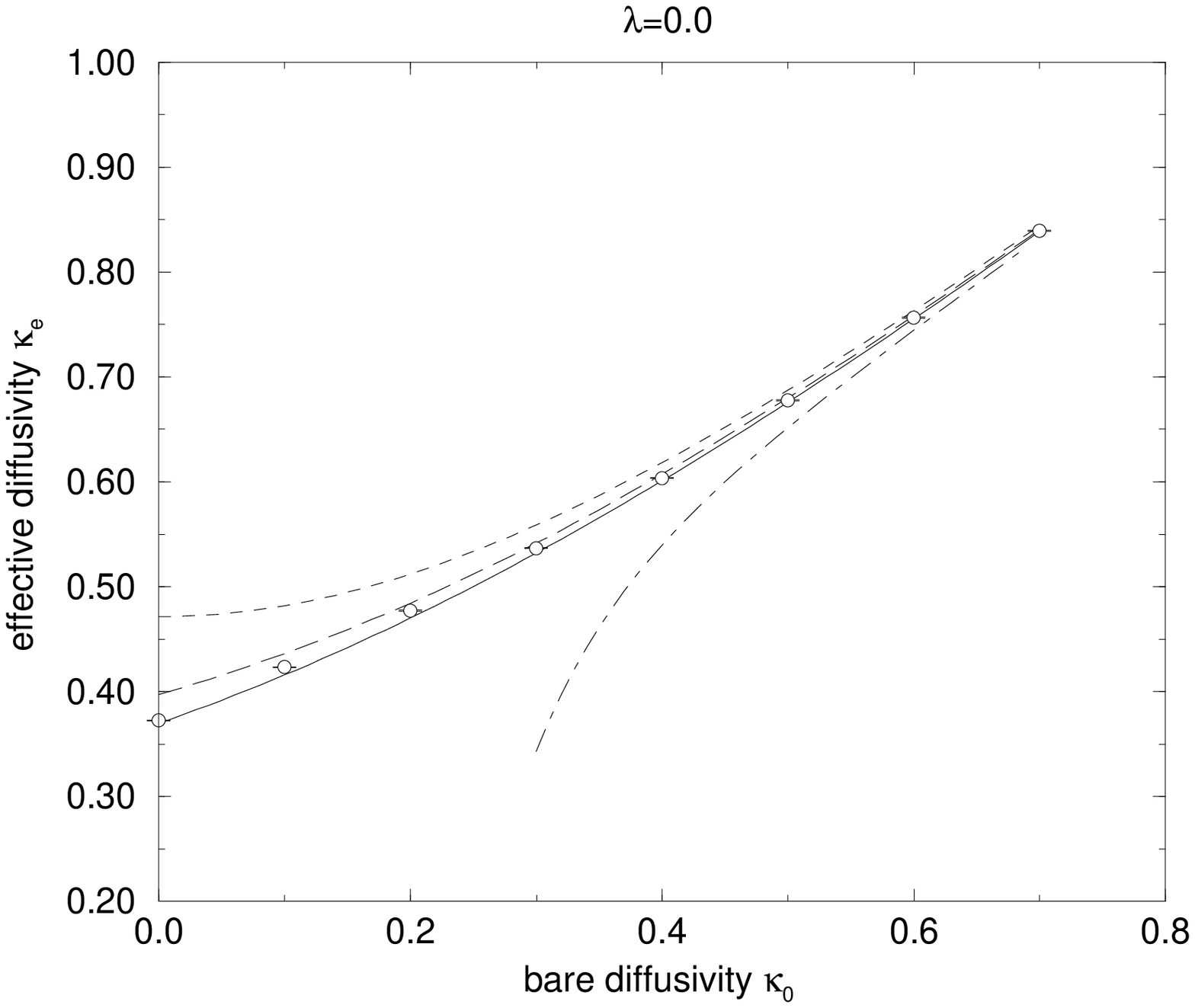} 
\end{center} 
\caption[]{\label{kevk000}\small $\k_e$ versus $\k0$ for fixed 
helicity $\l=0.0$. The simulation data are shown (\circo) to be 
compared with the predictions of two-loop self-consistent perturbation 
theory (solid), the Hartree-Fock calculation (long-dashed), the 
renormalization group (dashed) and ordinary perturbation theory 
(dot-dashed)} 
\end{figure}

\begin{figure}[h] 
\begin{center} 
\leavevmode \epsfxsize=14 truecm\epsfbox{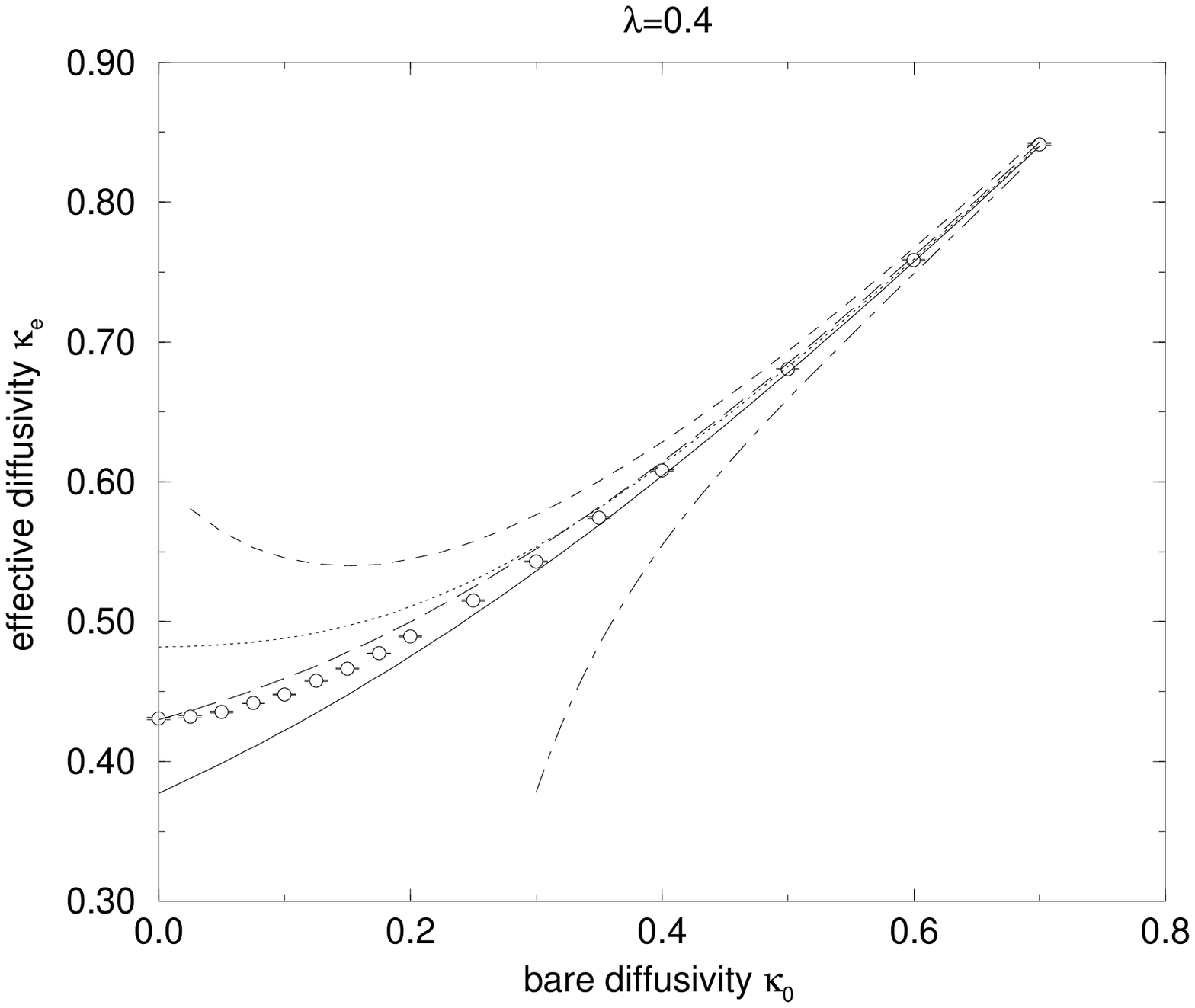} 
\end{center} 
\caption[]{\label{kevk004}\small $\k_e$ versus $\k0$ for fixed 
helicity $\l=0.4$. The simulation data are shown (\circo) to be 
compared with the predictions of two-loop self-consistent perturbation 
theory in $\k_e$ (solid) and in $\k_e,\gb$ (dotted), the Hartree-Fock 
calculation (long-dashed), the renormalization group (dashed), and 
ordinary perturbation theory (dot-dashed)} 
\end{figure}

\begin{figure}[h] 
\begin{center} 
\leavevmode \epsfxsize=14 truecm\epsfbox{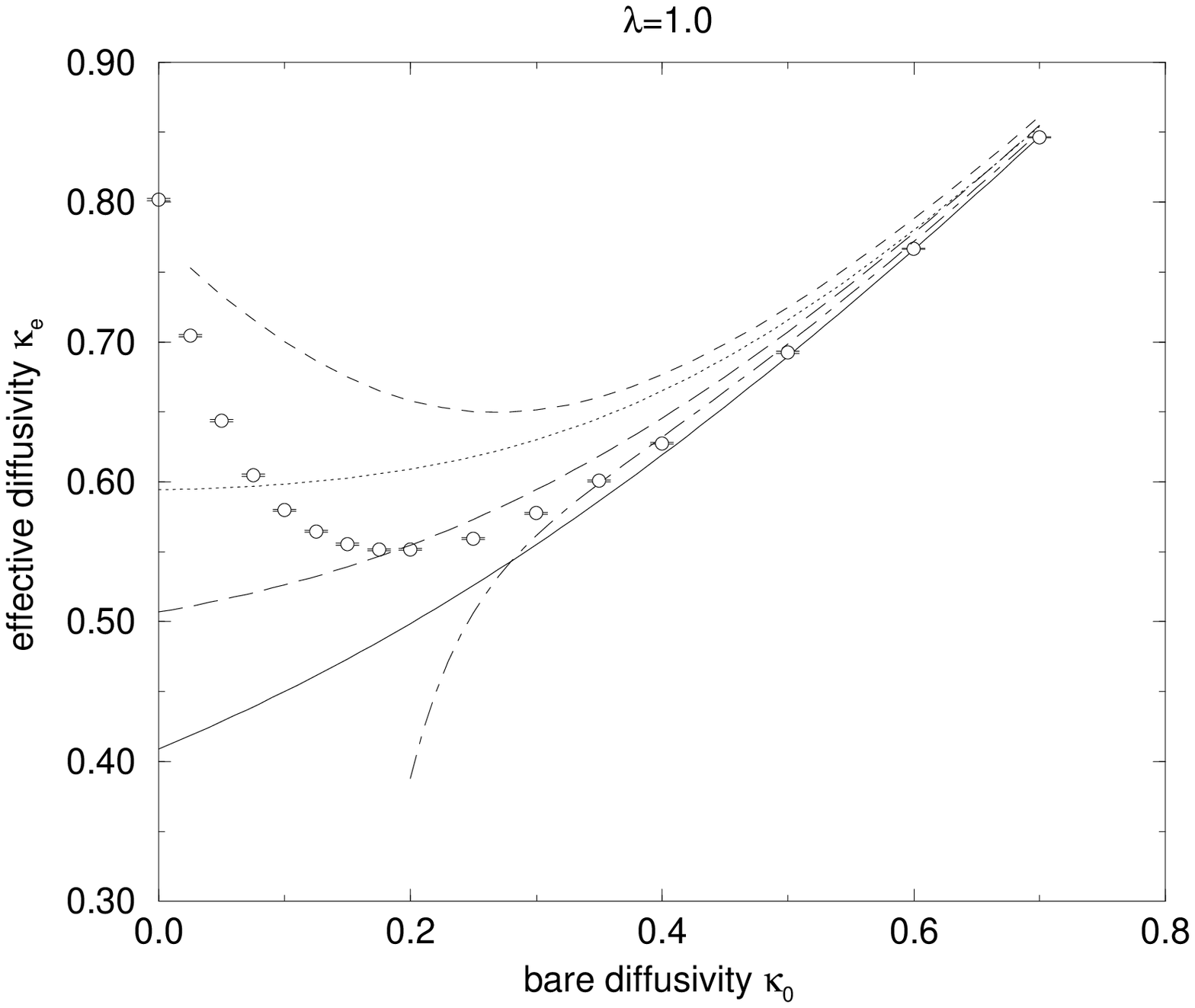} 
\end{center} 
\caption[]{\label{kevk010}\small $\k_e$ versus $\k0$ for fixed 
helicity $\l=1.0$. The simulation data are shown (\circo) to be 
compared with the predictions of two-loop self-consistent perturbation 
theory in $\k_e$ (solid) and in $\k_e,\gb$ (dotted), the Hartree-Fock 
calculation (long-dashed), the renormalization group (dashed), and 
ordinary perturbation theory (dot-dashed)} 
\end{figure}

\begin{figure}[h] 
\begin{center} 
\leavevmode \epsfxsize=14 truecm\epsfbox{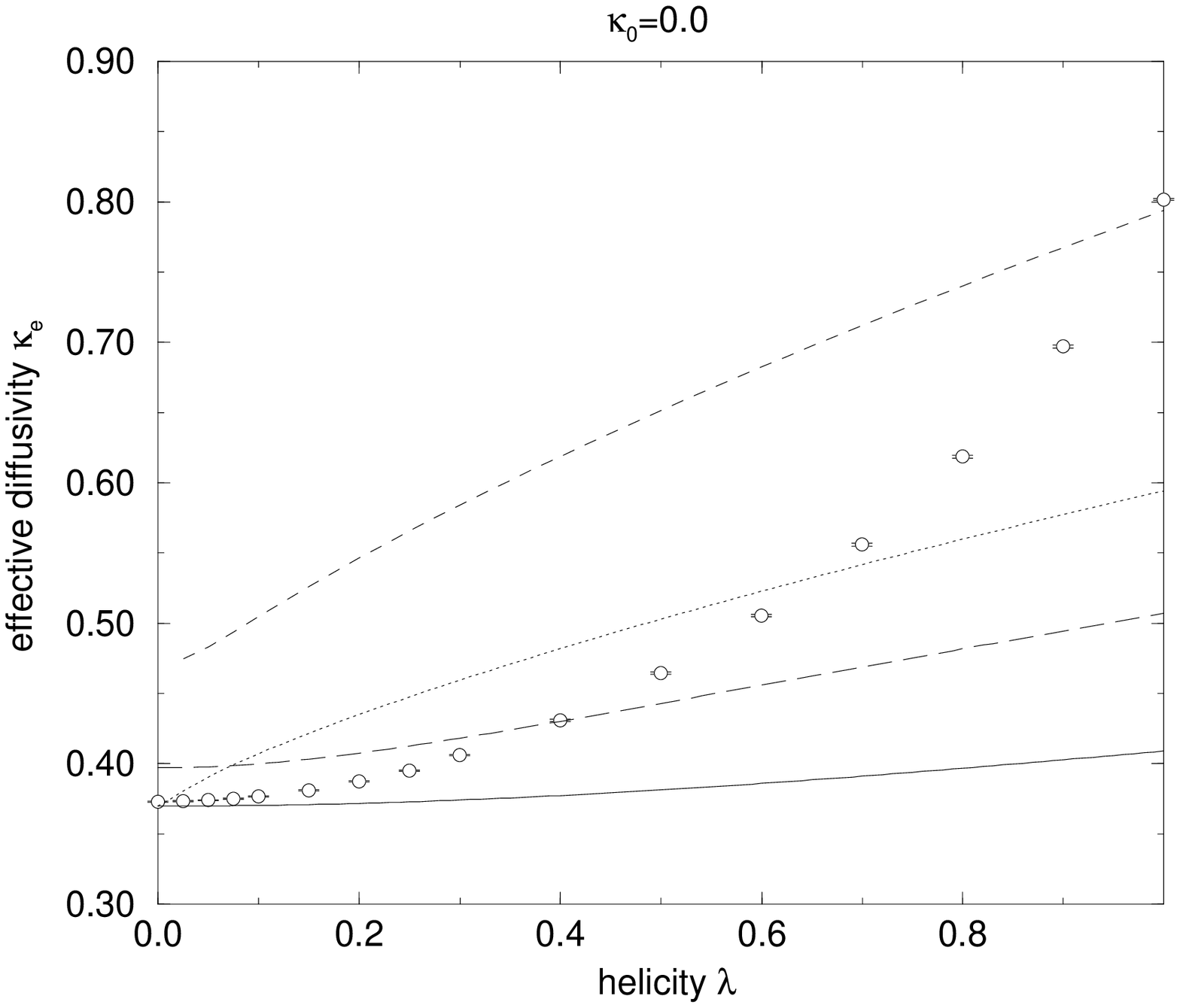} 
\end{center} 
\caption[]{\label{kevl00}\small $\k_e$ versus $\l$ for fixed 
diffusivity $\k_0=0.0$. The simulation data are shown (\circo) to be 
compared with the predictions of two-loop self-consistent perturbation 
theory in $\k_e$ (solid) and in $\k_e,\gb$ (dotted), the Hartree-Fock 
calculation (long-dashed), the renormalization group (dashed)} 
\end{figure}

\begin{figure}[h] 
\begin{center} 
\leavevmode \epsfxsize=14 truecm\epsfbox{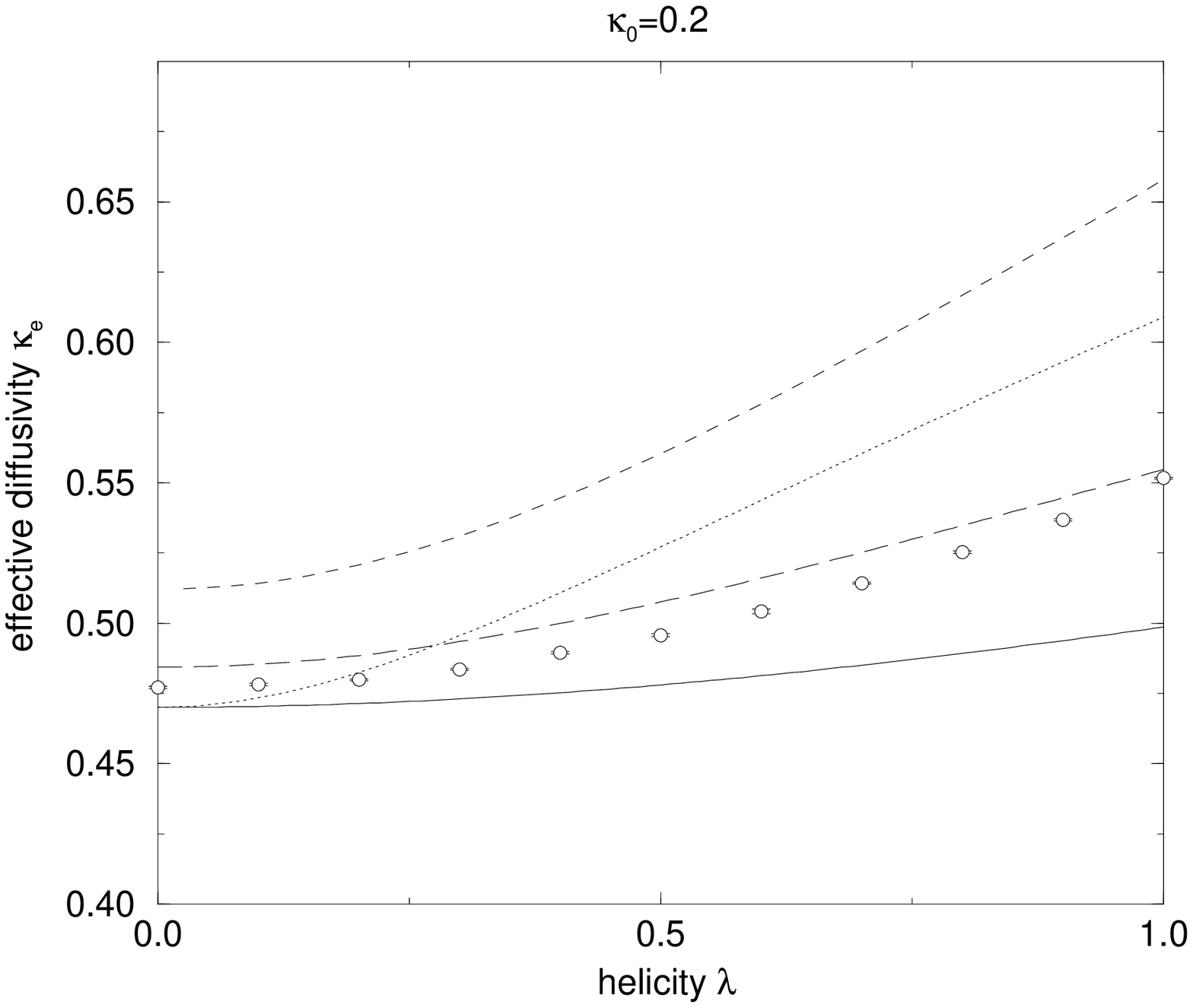} 
\end{center} 
\caption[]{\label{kevl02}\small $\k_e$ versus $\l$ for fixed 
diffusivity $\k_0=0.2$. The simulation data are shown (\circo) to be 
compared with the predictions of two-loop self-consistent perturbation 
theory in $\k_e$ (solid) and in $\k_e,\gb$ (dotted), the Hartree-Fock 
calculation (long-dashed), the renormalization group (dashed)} 
\end{figure}

\ben
\k_e~=~\k_0~+~\k_e\bb {1\over 9\k_e^2}~+~{0.0035~+~0.0059\l^2 
\over \k_e^4}\eb~.  \label{2LOOP_SC0} 
\een
We show the two-loop 
self-consistent prediction for $\k_e$ compared with data in figures 
\ref{kevk000} -- \ref{kevl02}. In figures \ref{kevk000}, \ref{kevk004} 
and \ref{kevk010} $~\k_e$ is plotted against $\k_0$ for fixed 
$l=0.0,0.4,1.0$ and in figures \ref{kevl00} and \ref{kevl02} $\k_e$ is 
plotted against $\l$ for fixed $\k_0=0.0,0.2$.  As should be expected, 
we see from figures \ref{kevk004} and \ref{kevk010} the agreement 
between theory and simulation is acceptable for $\k_0$ large enough. 
This is simply because the large molecular diffusivity swamps all 
other effects.  However, there is a large disagreement for small 
$\k_0$ which is most marked for $\k_0=0$.  The prediction for $\k_e$ 
behaves like $O(\l^2)$ and for $\k_0$ changes from $\k_e=0.3697$ at 
$\l=0$ to $\k_e=0.4090$ at $\l=1$: an increase of 10\%. In contrast, 
the simulation gives $\k_e=0.3705(1)$ and $\k_e=0.8018(7)$ 
respectively at these two values of $\l$: an increase of more than a 
factor of two. From the simulation for $\k_0$ small enough we find 
that $\k_e$ as a function of $\l$ is strongly in disagreement with the 
slow polynomial behaviour in $\l$ predicted by self-consistent 
perturbation theory. This effect was first observed in \c{DruDuaHor:1} 
and has remained unexplained. 
 
In addition, in figure \ref{kevk010} we observe a marked dip in the 
data at fixed $\l=1$ for $\k_e$ versus $\k_0$ at about $\k_0=0.2$. The 
major feature is that $\k_e$ rises rapidly with $\l$ at $\k_0=0$ 
whereas the effect for $\k_0 \ge 0.2$ is much less strong: the dip is 
not a lowering of the curve as $\l$ increases at $\k_0=0.2$ but rather 
a rapid rise with $\l$ at $\k_0=0$. The self-consistent prediction of 
this section does not predict a dip of any kind. 
 
\newsubsection{\label{SC_GEN} {\bf A more general approach}} 
 
In this section we propose an explanation of the enhancement of $\k_e$ 
by helicity in the flow. The technique is presented in detail at the 
one-loop level and the extension two-loop is then given. 
 
\begin{figure}[h] 
\begin{center} 
\leavevmode \epsfxsize=14 truecm\epsfbox{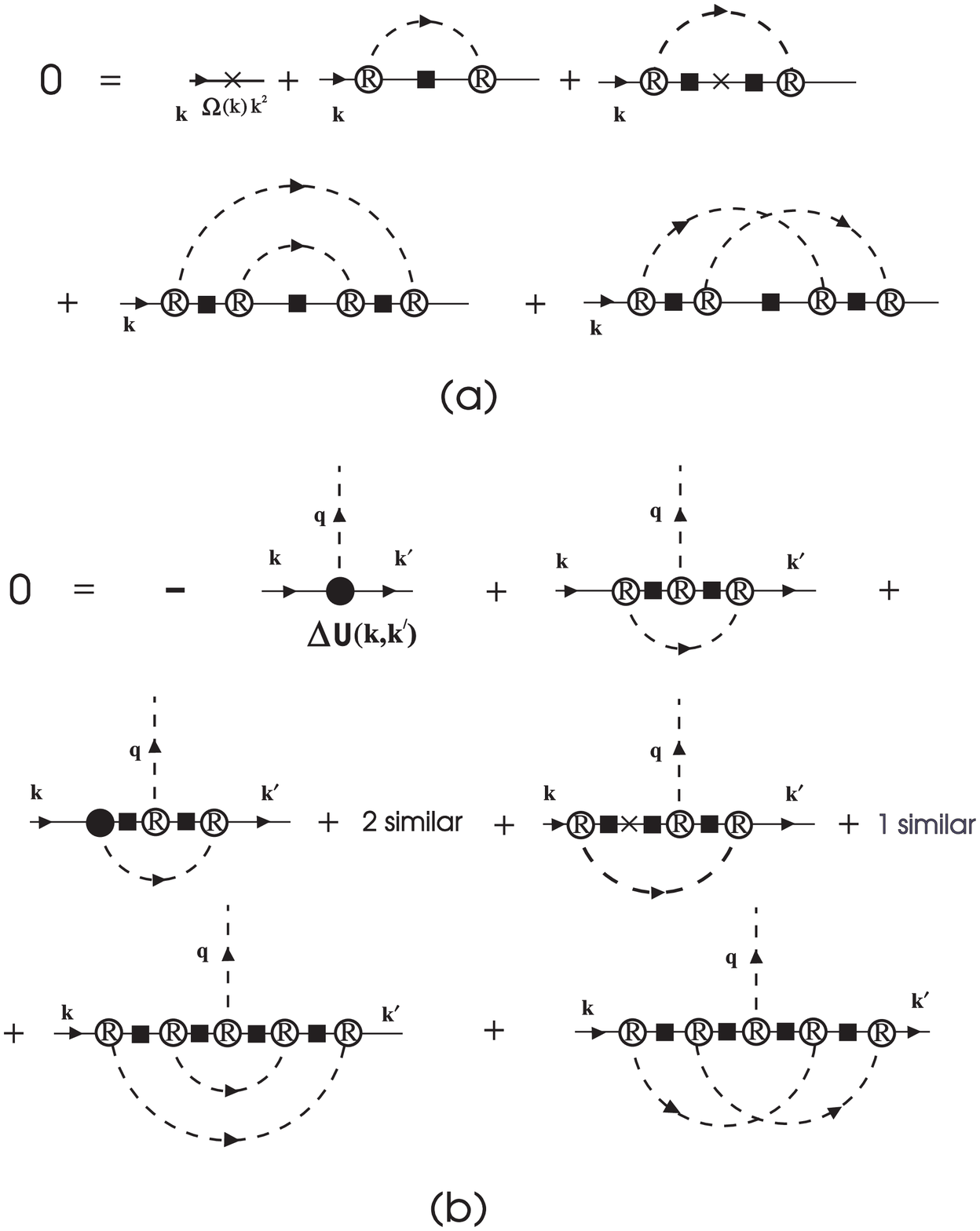} 
\end{center} 
\caption[]{\label{GEN_SC}\small The perturbation expansion to 
two-loops of the general self-consistent condition relating the 
self-energy, $\S(k)$, and vertex, $\U^R(k,\bkpr)$, functions. The 
vertex function is represented by the circle with inset 'R', full 
Green function, $\Gt(\bk)$, by the filled box and $\D\bU = 
\U^R-\U^0$.} 
\end{figure}

The philosophy is to suggest an effective, low wavevector diffusion 
equation obeyed by the smoothed distribution function. Because the 
wave vector is small it is assumed that the equation can be limited to 
at most two spatial derivatives.  The shortcomings of this assertion 
are discussed later. We propose the equation 
\ben
{\d\Th \over \d t} = 
\k_0 \nabla^2\Th - \a_R\,\bu\cdot\nabla\Th - 
\gb_R\,\bom\cdot\nabla\Th~,\label{RG_DIFFEQ} 
\een
where $\bom$ is the 
vorticity and $\a_R$ and $\gb_R$ are coupling constants which must be 
determined self-consistently. The bare values of these couplings, 
$\a_0$ and $\gb_0$, define the original diffusion equation. From 
eqn. (\ref{DIFFEQ}) we see that in our case $\a_0 = 1$ and $\gb_0 = 0$ 
but the analysis may be applied for general values of $\a_0$ and 
$\gb_0$. Eq. (\ref{RG_DIFFEQ}) is equivalent to renormalizing the 
velocity field to $\bu_R = \a_R\bu + \gb_R\bom$.  It will turn out 
that $\a$ is not renormalized and so the self-consistency conditions 
are applied only to determine $\k_e$ and $\gb_R$. To this end the 
equation for the $G(\bx)$ is taken to be 
\ben
\k_e \nabla^2 G - \D 
\k\: \nabla^2 G - \a_0\,\bu\cdot\nabla G - \gb_R\,\bom\cdot\nabla G + 
\D\gb\,\bom\cdot\nabla G ~=~-\de(\bx)~,\label{GREENEQ1} 
\een
where, as 
before, $\D \k = \k_e - \k_0~$ and $\D\gb = \gb_R - \gb_0~$. The rules 
for perturbation theory are the same as given in section \ref{PT} with 
the additional rules \bit 
\item[(vi)] The primitive vertex $\bW_i(\bkpr,\bk)$ associated with 
the vorticity and whose diagrammatic representation is shown in figure 
\ref{VERTEX}b, is given by $\bW_i(\bkpr,\bk) = (\bkpr \wedge \bk)_i$; 
\item[(vii)] For each vertex of type $\bV_i$ a factor of $\a_R$ and 
for each vertex of type $\bW_i$ a factor of $\gb_R$.  \eit The 
self-consistent equations are given by setting the next 
renormalizations of $\k_e$ and $\gb_R$ to zero in perturbation 
theory. This gives two equations which simultaneously determine $\k_e, 
\a_R$ and $\gb_R$ in terms of the bare parameters $\k_0, \a_0$ and 
$\gb_0$. It is convenient to define the general vertex 
$\bU_i(\bkpr,\bk)$ of the form 
\ben
\bU_i(\bkpr,\bk) = 
iV(\bkpr,\bk)\:\bkpr_i + W(\bkpr,\bk)\:(\bkpr \wedge \bk)_i~, 
\label{GEN_VERTEX} 
\een
where the form-factors $V$ and $W$ are scalar 
functions of $\bk$ and $\bkpr$.  The bare vertex $\bU_i^0$ is defined 
by $V^0 = \a_0,~W^0 = \gb_0$.  There is no independent form-factor 
coefficient proportional to $\bk_i$ in this expansion since the 
velocity field is incompressible. The diagrammatic representation of 
$\bU_i^0$ is shown in figures \ref{VERTEX}c and \ref{VERTEX}d, where 
the bare vertex is represented by an open circle while the 
renormalized vertex carries additionally an inset letter 
`R'. Likewise, the general expression for $\la \Gt(\bk) \ra$ can be 
defined as 
\ben
\la \Gt(\bk) \ra~=~{1 \over \bk^2\bb\k_0 + 
\Om(\bk^2)\eb}\label{OMEGA} 
\een
To two loops the self-consistent 
relationships that hold between diagrams are shown in figure 
\ref{GEN_SC}. 
 
We are interested in the low-wavenumber properties of the theory and 
we use the approximations \bea \Om(\bk^2)&=&(\k_e - 
\k_0)~+~O(\bk^2)~,\nn\\ 
V^R&=&\a_R~+~O(\bk^2,\bkpr^2,\bk\cdot\bkpr)~,\nn\\ 
W^R&=&\gb_R~+~O(\bk^2,\bkpr^2,\bk\cdot\bkpr)~. \label{LOWK} \eea These 
approximations are consistent with the form considered for the 
effective equation governing the dispersal, eqn. (\ref{RG_DIFFEQ}), 
since the renormalized couplings, $\a_R$ and $\gb_R$, are given 
respectively by the coefficients of $\bkpr_i$ and $(\bkpr \wedge 
\bk)_i$ in the expansion of $\bU_i$. To include higher powers of $\bk$ 
and $\bkpr$ would, for consistency, require terms involving $\bu$ and 
$\bom$ in eqn. (\ref{RG_DIFFEQ}) with higher powers of $\nabla$. In 
principle, a functional self-consistent formalism could then be set up 
for $\Om(\bk)$ and $\bU_i(\bkpr,\bk)$ treating them respectively as a 
vector and matrices in wave-number space. However, this is a further 
generalization that we have not yet pursued because the amount of 
computing effort required to determine them numerically is 
prohibitive.  In the present case we are interested only in the 
low-wavevector behaviour of these functions, and then the 
self-consistent equations are determined by asserting the 
self-consistency in the limit $\bk^\prime,\bk \to 0$ and using 
eqns. (\ref{LOWK}). 
 
\begin{figure}[h] 
\begin{center} 
\leavevmode \epsfxsize=8 truecm\epsfbox{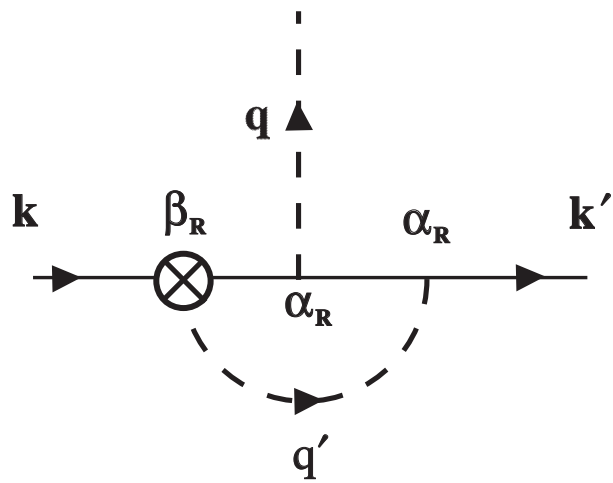} 
\end{center} 
\caption[]{\label{A_RENORM}\small An example of the kind of vertex 
graph that must be evaluated in the solution of equations shown in 
figure \ref{GEN_SC} once the approximation for $\bU^R$ given in 
eqn. (\ref{LOWK}) and shown in figure \ref{VERTEX} has been used. This 
graph is labelled $T_{\gb\a\a}~$.} 
\end{figure}

The calculation is now straightforward but tedious. All loop integrals 
are approximated by their lowest non-zero power in $(\bk,\bkpr)$ and 
their contribution to the renormalization of the relevant coupling 
constant is read off. It suffices to give some examples which indicate 
how the full result is obtained and to this end we first analyze in 
detail the one loop approximation to the self-consistent 
equations. The first observation is that $\a$ is not renormalized, 
i.e., $\a_R = \a_0$. We give one example indicating how this comes 
about.  The graphs is labelled by the couplings corresponding to the 
types of vertex it contains and the label is ordered in the same order 
that they occur in the graph.  We consider the contribution shown in 
figure \ref{A_RENORM} to the vertex renormalization and will 
concentrate on the part proportional to $\bkpr$.  The value of this 
graph is 
\ben
T_{\gb\a\a} = \a_R^2\gb_R\int \dq 
{\e_{lmn}\bk_m\bq_n\,(\bkpr-\bq)_p\bkpr_i\,F_{lp}(\bq) \over 
\k_e(\bk-\bq)^2\:\k_e(\bkpr-\bq)^2}~.\label{TBAA} 
\een
The 
approximations of eqn. (\ref{LOWK}) have been implemented.  Only the 
helical part of $F_{ij}(\bq)$ contributes and we find the result \bea 
T_{\gb\a\a}&=& {\a_R^2\gb_R\l \over \k_e^2}\,i\bkpr_i\int \dq 
{\e_{lmn}\bk_m\bq_n\,(\bkpr-\bq)_p q\,\e_{lpq}\bq_q\,\Phi(\bq) \over 
(\bk-\bq)^2(\bkpr-\bq)^2} \nn\\ &=&{\a_R^2\gb_R\l \over 
\k_e^2}\,i\bkpr_i\int \dq {\bb \bk\cdot\bkpr\,\bq^2 - 
\bk\cdot\bq\,\bkpr\cdot\bq \eb q \Phi(\bq) \over 
(\bk-\bq)^2(\bkpr-\bq)^2}~.  \label{TBAA0} \eea Clearly the 
contribution to $V$ is $O(\bk\cdot\bkpr)$ and so $\a$ is not 
renormalized. All contributions to $V$ are similarly of higher order 
and the result is that $\a_R = \a_0$. 
 
The coupling $\gb$ is renormalized when $\l \neq 0$. The calculation 
follows a similar path to that used in the analysis of the 
renormalization of $\a$.  Again we show one calculation explicitly and 
consider the contribution to $W(\bk,\bkpr)$ by calculating the 
coefficient of $(\bkpr \wedge \bk)$ in $T_{\a\a\a}$: 
\ben
T_{\a\a\a}~=~-i{\a_R^3 \over \k_e^2}\int \dq 
{(\bk-\bq)_l(\bkpr-\bq)_n\,F_{ln}(\bq) \over 
(\bk-\bq)^2(\bkpr-\bq)^2}~.\label{TAAA} 
\een
The contribution 
proportional to $(\bkpr \wedge \bk)$ comes only from the helical part 
of $F_{ln}(\bq)$ and so the relevant term is 
\ben
T_{\a\a\a}~\sim~{\a_R^3\l \over \k_e^2}\int \dq 
{\e_{lnp}\bk_l\bkpr_n\bq_p\bq_m\,q\Phi(\bq) \over 
(\bk-\bq)^2(\bkpr-\bq)^2}~.\label{TAAA0} 
\een
Hence we find the 
contribution $\de\gb_R$ to the renormalization of $\gb$ from 
$T_{\a\a\a}$ to be 
\ben
\de\gb_R~=~-{\a^3\l \over 6\pi^3\k_e^2}\int \d 
q\:q\Phi(\bq)~.\label{DBAAA} 
\een
The renormalization of $\gb_R$ is 
expressed in terms of three integrals: 
\ben
I_n~=~{1 \over 6\pi^3}\int 
\d q\:q^n\Phi(\bq)~~~~~~n~=~1,2,3~. \label{INTS} 
\een
After evaluating 
all the relevant graphs the self-consistent equations are \bea 
\a_R-\a_0&=&0~, \nn\\ 
\gb_0-\gb_R&+&(\a_R^3\l\,I_1~+~2\a_R^2\gb_R\,I_2~+~\a_R\gb_R^2\l\,I_3)~=~0~. 
\label{ABRENORM} 
\eea 
 
The approximate equation for $\Gt(\bk)$ is given by the equation for 
$\S(\bk)$ in terms of the one-particle irreducible graphs in figure 
\ref{GEN_SC} at one-loop order. Because we are using the 
low-wavenumber approximation this reduces to substituting the 
expression for the renormalized vertex $\bU_i(\bkpr,\bk)$ given in 
eqns.  (\ref{GEN_VERTEX}) and (\ref{ABRENORM}) into the one-loop 
diagram for $\S(\bk)$ in figure \ref{GEN_SC}. We analyze the one-loop 
self-energy graph and keep only the term proportional to $\bk^2$.  In 
obvious notation this gives the results 
\ben
T_{\a\a}~\sim~-{2\a_R^2 
\over \k_e}\,I_2~~~~~~~~T_{\gb\a}~\sim~ {2\a_R\gb_R\l \over 
\k_e}\,I_3~.\label{SELF} 
\een
Using the spectra in 
eqn. (\ref{SPECTRA}) eqns. (\ref{ABRENORM}) and (\ref{SELF}) for the 
one-loop self-consistent conditions become: 
\ben
\D\a~\equiv~\a_R-\a_0~=~0~,~~-\D\gb~+~{B_1 \over \k_e^2}~=~0~,~~ 
\D\k~+~{C_1 \over \k_e}~=~0~, \label{ALLRENORM} 
\een
where \bea 
B_1&=&{1 \over 18}\bb 2\a_R^2\gb_R + 
2\sqrt{2\over\pi}\:\a_R\gb_R^2\,\l + \sqrt{2\over\pi}\:\a_R^3\l 
\eb~,\nn\\ C_1&=&-{1 \over 9}\bb \a_R^2 + 
4\sqrt{2\over\pi}\:\a_R\gb_R\,\l + 3\:\gb_R^2 \eb~. \label{B1C1} \eea 
>From these equations it is clear that no renormalization occurs if 
there is no pseudo-scalar or axial-vector quantity in the problem: if 
$\gb_0 = \l = 0$ then the problem reduces to the one-loop 
self-consistent analysis presented in section \ref{SCke}. However, if 
either $\gb_0$ or $\l$ are non-zero then $\gb$ is renormalized and the 
effect on $\k_e$ is encoded in eqn. (\ref{ALLRENORM}).  In our case we 
set $\a_R = \a_0 = 1,~\gb_0 = 0$ and $\l \neq 0$. The equations 
(\ref{ALLRENORM}), (\ref{B1C1}) then give \bea &&3\sqrt{\pi\over 
2}\:\gb^3~+~3\:\gb_R^2\l~+~9\sqrt{\pi\over 2}\:\k_e\k_0\gb_R~-~ {\l 
\over 2}~=~0~, \nn\\ &&\k_e~=~\k_0~+~{1 \over 9\k_e}\bb \a_R^2 + 
4\sqrt{2\over\pi}\:\a_R\gb_R\,\l + 3\:\gb_R^2 \eb~. \label{CUBE} \eea 
For small $\l$ and $\k_0=0$ we deduce that 
\ben
\gb \sim \bb {1\over 
18\pi}\eb^{1/6}\,\l^{1/3}~\Rightarrow~ \k_e~\sim~{1 \over 3}~+~{1 
\over 2}\bb {1\over 18\pi}\eb^{1/3}\,\l^{2/3}~.\label{SMALL_L} 
\een

The data for $\k_e$ versus $\l$ for $\k_0=0$ is shown in figure 
\ref{kevl00} and we see that for small $\l$ the simulation results are 
not compatible with $\l^{2/3}$ behaviour. We shall see below that this 
is not rectified in the two-loop self-consistent calculation. However, 
in this one-loop calculation there is a considerable enhancement in 
the dependence of $\k_e$ on $\l$, whereas in the self-consistent 
calculation of section \ref{SCke}, in which the generation of the new 
vertex coupled to the vorticity $\bom$ was not included, there is no 
effect at all at one-loop order and only a mild effect at two-loop 
order. The equations (\ref{CUBE}) can be solved numerically.  For 
example, for $\k_0=0,~\l=1$ we find $\gb_R=0.3456$ and the effective 
velocity field is predicted to be 
\ben
\bu_R~=~\bu~+~\gb_R\,\bom~,\label{EFF_V} 
\een
which clearly leads to 
an enhanced effective diffusivity, $\k_e = 0.5207$, compared with 
$\k_e = 0.4090$ from the two-loop calculation of the previous 
section. We believe that we have qualitatively captured the mechanism 
responsible for the enhancement of the effective diffusivity by 
helicity. 
 
The one-loop calculation is limited because it is not accurate at 
$\l=0$ unlike the two-loop calculation. We have investigated the 
two-loop extension of the self-consistent approach when the new vertex 
with coupling $\gb$ is included. This is more involved and the 
integrals were done numerically. We present the final results below. 
 
The two-loop self-consistent equations are \bea \D\k + {C_1 \over 
\k_e} -\D\gb {1\over \k_e}{\pd C_1\over \pd \gb} + {1\over 
\k_e^3}(C_2-C_1^2)&=&0~,\nn\\ -\D\gb + {B_1\over \k_e^2} + 2{\D\k 
\over \k_e^3}B_1 - \D\gb{1\over \k_e^2}{\pd B_1\over \pd \gb} + 
{B_2\over \k_e^4}&=&0~, \label{SC2} \eea where $B_1$ and $C_1$ are 
given in eqn. (\ref{B1C1}) and $B_2$ and $C_2$ are evaluated 
numerically to be ($\a_R=\a_0=1$) \bea 
B_2&=&-(0.0047\l+0.0095\gb+0.0180\gb\l^2+0.0644\gb^2\l+0.0423\gb^3+\nn\\ 
&&0.0252\gb^3\l^2+0.0287\gb^4\l)~,\nn\\ 
C_2&=&0.0088-0.0034\l^2-0.0165\gb\l-0.0315\gb^2-0.0110\gb^2\l^2-\nn\\ 
&&0.0514\gb^3\l+0.0090\gb^4-0.0337\gb^4\l^2~. \label{B2C2} \eea 
Eqns. (\ref{SC2}) can be rearranged to give \bea \D\k + {C_1 \over 
\k_e^2} + {1 \over \k_e^3}(C_2-C_1^2-B_1{\pd C_1 \over 
\pd\gb})&=&0~,\nn\\ -\D\gb + {B_1 \over \k_e^2} + {1\over 
\k_e^4}(B_2-2B_1C_1-B_1{\pd B_1 \over \pd\gb})&=&0~.\label{SC2_0} \eea 
We shall set $\gb_0=0$ from now on. These equations contain the 
accurate two-loop self-consistent result which fits the data for all 
$\k_0$ at $\l=0$.  For $\l \ne 0$ these equations are solved 
numerically and the results are compared with simulation data in 
figure \ref{kevk000} -- \ref{kevl02} As in the one-loop case the 
behaviour for small $\l$ is clearly incorrect and there is no 
quantitative agreement with the data. However, there is a clear 
enhancement of $\k_e$ due to the inclusion of the vorticity vertex and 
associated coupling $\gb$ and for $\l=1.0$ we find $\k_e=0.5959$. 
This mechanism for enhancing turbulent diffusion cannot obviously be 
deduced from perturbative considerations. It arises from a resummation 
of diagrams which give an expression that analytically continues 
between the regions where $\l^2 \ll u_0^2/\k_0^4 k_0^4$ and $\l^2 \gg 
u_0^2/\k_0^4 k_0^4$. The effective diffusivity in the former region is 
well predicted by perturbation theory, but not so for parameters in 
the latter region.  Although quantitative agreement is not good the 
important point is that a mechanism has been discovered which gives a 
strong enhancement to the value of $\k_e$ for non-zero $\l$ even in 
the one-loop approximation, whereas in the self-consistent theory for 
$\k_e$ alone, discussed in the previous section, there is no effect of 
helicity at all on the value of $\k_e$ at one-loop order.  The obvious 
reason for the discrepancy in this new approach is that the 
approximations made are much too crude. A more refined calculation 
would use a functional self-consistent method for $\bU_i(\bkpr,\bk)$ 
(eqn. (\ref{GEN_VERTEX})) and $\Om(k^2)$ 
(eqn. (\ref{OMEGA})). Although a computationally formidable task, this 
is likely to encode the correct behaviour much more accurately than 
does our low-wavenumber approximation. 
 
The origin of the dip in figure \ref{kevk010} in the curve $\k_e$ 
versus $\k_0$ for $\l=1$ is unexplained by the theory presented so 
far. 
 
\newsection{\label{HFOCK} The Functional Hartree-Fock Method} 
 
This approach goes some way towards including effects omitted in the 
low-wavenumber approximation. The version presented here is deficient 
in that the prediction for $\k_e$ when $\k_0=\l=0$ is not as accurate 
as the two-loop self-consistent approach but the advantage is that 
$\Om(k^2)$, eqn. (\ref{OMEGA}), is treated as function to be 
determined self-consistently by the Hartree-Fock equations.  The 
vertices are still treated in the low-wavenumber approximation and, as 
in the previous section, they are parameterized by $\a$ and $\gb$. 
 
\begin{figure}[h] 
\begin{center} 
\leavevmode \epsfxsize=14 truecm\epsfbox{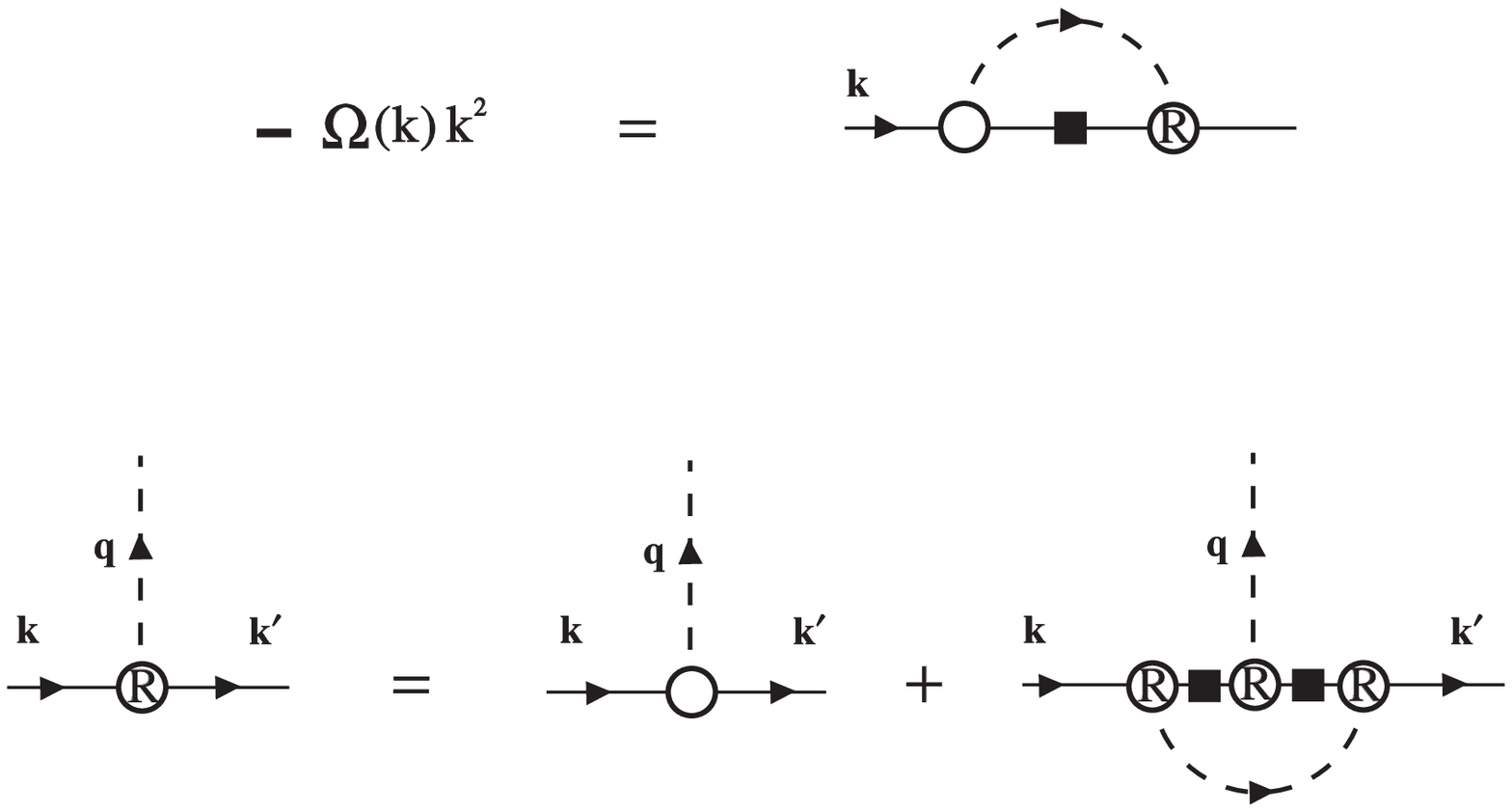} 
\end{center} 
\caption[]{\label{HF}\small The integral Hartree-Fock equations for 
$\O(\bk)$ and $\U^R~$ in terms of the general vertices $\bU^0$ and 
$\U^R~$. The full Green function, $\Gt(\bk)$, is denoted by the filled 
box. The equation for $\O(\bk)$ is correct to all loop orders but the 
equation for $\U^R~$ is correct to one-loop order only.} 
\end{figure}

The integral equation to be satisfied by $\Om(k^2)$ and the one-loop 
equation satisfied by the vertex function, which is the same as the 
one-loop self-consistent equation, are shown in figure \ref{HF}. Note 
that, unlike the self-consistent calculation of the previous section, 
only one of the vertices in the one-loop self-energy is replaced with 
the full vertex since this gives the correct counting of diagrams when 
the equations are iterated. The self-consistent case is different 
because the augmented vertex is already present in the perturbation 
theory and corrections are implemented by counter-terms.  The 
approximation for the vertices in eqn. (\ref{LOWK}) is used and $\gb$ 
is determined using eqn. (\ref{ALLRENORM}): 
\ben
\gb~=~\gb_0~+~{B_1\over \k_e^2}~, \label{BETA_HF} 
\een
where $B_1$ is 
given in eqn. (\ref{B1C1}) and using $\k_e = \k_0 + \Om(0)~$.  The 
Hartree-Fock equation to be satisfied by $\Om(k^2)$ is then (after 
some reduction and using $\phi$ in eqn. (\ref{SPECTRA})) \bea 
\Om(k^2)&=& {1\over k^2}\left[ {2\over 3\sqrt{2\pi}}\;e^{-k^2/2}\int 
dp\;{\left(pk\;\cosh(pk)- \sinh(pk)\right)e^{-p^2/2}\over 
pk\;(\k_0+\Om(p^2))}~+~ \right. \nn\\ &&\left. 
~~~~\gb\l\;\int{d^3p\over(2\pi)^3}\:{|\bp+\bk|\;\phi(|\bp+\bk|)\; 
(k^2p^2-(\bk\cdot\bp)^2) \over \bp^2(\k_0+\Om(p^2))} \right] \eea 
These equations are solved by discretizing the wavevector, performing 
the integrals numerically and iterating the equations to converge to a 
solution for $\Om(k^2)$. The effective diffusivity is then $\k_e = 
\k_0 + \Om(0)~$. 
 
The results are shown in figures \ref{kevk000} -- \ref{kevl02}, where 
it is clear that while the numerical value predicted at $\k_0=\l=0$ 
is not accurate, the behaviour for small $\l$ is more in keeping with 
the simulation results. This strengthens our belief that an analysis 
which treats the propagator and vertex as functions to be determined 
self-consistently is likely to reproduce the desired properties more 
accurately. However, at $\k_0=0, \l=1$ the value predicted is 
$\k_e=0.5070$, still much less than the simulation value of 
$0.8018(7)$. 
 
In principle, the vertex may be treated as a function in the same 
manner as $\Om(k^2)$. This is prohibitively expensive in memory and 
computer time but might be possible if some simplification of the 
function form were implemented.  Also, while the equation for 
$\Om(k^2)$ is already exact at one-loop, that for the vertex is not 
and we cannot preclude that higher loop corrections might be 
important. We have not pursued this approach. 
 
We note that in this approach, as with those of the previous sections, 
  the marked dip in $\k_e$ as a function of $\k_0$ for the larger 
  values of $\l$ is not reproduced.

\newsection{\label{RG} \bf The Renormalization Group} 
 
In the previous section we presented an analysis based on the 
assumption that the large-scale advection is controlled by an 
effective transport equation dominated by the terms containing only 
one and two derivatives. This method is related to the renormalization 
group (RG) methods which have proved very successful in predicting 
exponents in critical phenomenon. In the RG approach a large 
wavenumber cutoff, $\L$, is introduced and the advection on scales 
larger than $L \equiv 2\pi/\L$ is assumed to be described by an 
effective transport equation, in principle containing terms with an 
arbitrarily high number of derivatives.  The parameters in this 
equation are functions of $\L$ in order to account for the effect of 
advection at the scales smaller than $L$ which have been excised.  In 
the limit $\L \rightarrow 0$ the effective equation, by dimensional 
analysis, takes a simple form dominated by terms with few derivatives 
and with associated effective or ``renormalized'' parameters. In this 
way the effective equation takes a form similar to that used in the 
previous section. There is a difference, though, because any practical 
application of these schemes requires a drastic truncation of the 
operator space: especially in the RG method where it is impossible to 
compute the flow with changing $\L$ for very many parameters in the 
effective transport equation. Unlike the situation in critical 
phenomena there are no infra-red divergences in the theory and the 
notion of a relevant operator is not applicable. It is then a matter 
of trial and error to determine whether the approach used captures the 
vital features controlling the flow. The simplest renormalization 
scheme is to calculate the renormalization to the diffusivity $\k(\L)$ 
and to the vertex associated with the coupling of the random field or 
externally applied drift. ln the case of gradient flows we 
demonstrated in reference \c{DeaDruHor:2} that this scheme yields 
exact results in one and two dimensions and an extremely accurate, 
although not exact, result in three dimensions. It is, in general, 
much harder to calculated the renormalized parameters such as $\k_e$ 
than the associated exponents, and so success in \c{DeaDruHor:2} 
suggests that some insight may be gained using RG methods in other 
similar problems. 
 
In this section we present a RG calculation of $\k_e$. The vertex 
renormalization is done but multiple vertex renormalization is 
neglected which means that the renormalized velocity field remains 
Gaussian. Consequently, after integrating out the random field down to 
wave number $\L$ we postulate that the equation for the effective 
Green function can be approximated, for all $\L$, by an equation of 
the same form as the original one (eqn. \ref{GREENEQ}): 
\ben
\k(\L)\:\nabla^2 G(\bx,\L)~-~\bu(\bx,\L)\cdot\nabla G(\bx,\L) = 
-\de(\bx)~, \label{GEQ_RG} 
\een
where $\k(\L)$ is the running 
renormalized diffusion constant and $\bu_\L$ is the renormalized 
velocity field. Since we renormalize the vertex functionally the field 
correlation function will flow under the RG as 
\ben
\la 
\u_i(\bk,\L)~\u_j(\bk^\prime,\L) \ra~=~ \left\{~~ \ba{cl} 
(2\pi)^3\de(\bk+\bk^\prime)\;F_{ij}(\bk,\L)& |\bk|~~<~~\L \\ 0 & 
|\bk|~~>~~\L \\ \ea \right. \label{UU_RG} 
\een
where 
\ben
F_{ij}(\bk,\L)~=~\Phi(\bk,\L)(k^2\de_{ij} - 
k_ik_j)~+~\Psi(\bk,\L)i\e_{imj}k_m~.  \label{F_RG} 
\een
One finds that 
the renormalized field is still incompressible. We shall compute the 
flow equations for $\k(\L),~\Phi(\bk,\L)$ and $\Psi(\bk,\L)$ as $\L$ 
varies. 
 
The change in $\k(\Lambda)$ on integrating out wave vectors in the 
shell $(\Lambda, \Lambda - \delta \Lambda)$ is 
 
\begin{equation} 
\delta \k(\Lambda) = -{1\over 3 \pi^2\k(\Lambda)} \Lambda^2 
\Phi(\Lambda,\Lambda) \delta \lambda 
\end{equation} 
 
If one calculates the vertex renormalization and treats it as an 
addition to the random field one finds, 
\begin{equation} 
i\bk\cdot \delta {\ut}(\bq,\Lambda) + O(k^2) = -{i\over 
(2\pi)^3}\int_{\Lambda - \delta \Lambda}^\Lambda {F_{ij}(\bq^\prime, 
\Lambda) k_i q_j q^\prime_k {\u}_k(\bq, \Lambda) \over \k^2(\Lambda) 
q^{\prime 2} (\bq + \bq^\prime)^2} d{\bq^\prime} 
\end{equation} 
where we have only kept the vertex term to lowest order in $k$ as it 
is only this term that contributes to the one loop diffusivity 
renormalization. In addition if one assumes that it is the low wave 
number (long distance) renormalization of the velocity field which is 
important for the effective diffusivity one finds to lowest order in 
$q$, 
 
\begin{equation} 
\delta {\u}_i(\bq, \Lambda) \approx -{1\over (2\pi)^3\k^2(\Lambda)}q_j 
{\u}_k(\bq, \Lambda) \int_{\Lambda - \delta \Lambda}^\Lambda 
{F_{ij}(\bq^\prime \Lambda)q^\prime_k \over q^{\prime 4}} d\bq^\prime 
\end{equation} 
In the case where no helicity is present we see that the vertex is not 
renormalized. However (as pointed out in the section on perturbation 
theory) whilst the helicity does not contribute to the diffusivity 
renormalization at a one loop level it renormalizes the velocity 
field. The renormalization is zero at order $0$ in $q$ but has an 
order $1$ effect: 
\begin{equation} 
\delta {\u}_i(\bq, \Lambda) \approx -{1\over 6 \pi^2\k^2(\L)} 
\epsilon_{ijk} q_j \u_i(\bq, \Lambda) \Psi(\Lambda, \Lambda) \delta 
\Lambda 
\end{equation} 
In real space therefore, the renormalization is of the form 
\ben
\bu 
\to \bu + \delta\Lambda \gb(\Lambda) \nabla \times \bu.  
\een
Using 
the renormalization of $\ut$ one may compute the flow of $F_{ij}$ and 
thus $\Phi$ and $\Psi$ to obtain the one loop functional RG equations: 
\begin{eqnarray} 
{\partial \k \over \partial \Lambda} &=& -{1\over 3\pi^2 
 \k(\Lambda)}\Lambda^2 \Phi(\Lambda,\Lambda) \cr {\partial 
 \Phi(q,\Lambda) \over \partial \Lambda} &=& - {1\over 3\pi^2 
 \k^2(\Lambda)} \Psi(q,\Lambda)\Psi(\Lambda,\Lambda) \cr {\partial 
 \Psi(q,\Lambda) \over \partial \Lambda} &=& - {1\over 3\pi^2 
 \k^2(\Lambda)} q^2 \Phi(q,\Lambda)\Psi(\Lambda,\Lambda) 
\label{eq:rgf} 
\end{eqnarray} 
The integration of the eqns. (\ref{eq:rgf}) is from $\Lambda = \infty$ 
to $0$ with the initial conditions 
\begin{eqnarray} 
\k(\infty) &=& \k_0 \cr \Phi(q,\infty) &=& \Phi(q) \cr \Psi(q,\infty) 
&=& \Psi(q). 
\end{eqnarray} 
 
When there is no helicity there is no vertex renormalization at the 
order we are considering and therefore we may integrate the equations 
directly to obtain 
\begin{equation} 
\k_e = ( \k_0^2 + {2 u_0^2/9})^{1\over 2}~.\label{RG_L=0} 
\end{equation} 
This agrees with the one loop perturbation result, as it should, and 
in the case $\k_0 = 0$ we find that $\k_e = \sqrt{2}/3 = 0.47140$ 
which is quantitatively not very close to the numerically measured 
result, $\k_e = 0.3697$.  However, the discrepancy is sensitive to the 
form assumed for the effective diffusion equation. In our case this is 
given by eqn. (\ref{GEQ_RG}) which is clearly inadequate since $\bu$ 
is not renormalized when $\l=0$. An improvement can only be made by 
including terms with higher derivatives of $\bu$. This is similar to 
parameterizing the non-helical form factor $V^R$ of 
eqn. (\ref{GEN_VERTEX}) with a function of external momenta rather 
than approximating it by a constant, $\a_R$, which is not 
renormalized.  This is a possible avenue of research but we have not 
yet followed it. 
 
In contrast, for $\l \ne 0$, $\bu$ is renormalized and the effect on 
$\k_e$ is significant because the helical form factor $W^R$, 
eqn. (\ref{GEN_VERTEX}) is renormalized at low wavenumber as 
parametrized by $\gb(\L)$ above. The RG equations may be integrated 
numerically and is compared with simulation in figures \ref{kevk000} 
-- \ref{kevl02}. Although the results are not quantitatively accurate, 
they capture the qualitative behaviour seen in the simulations. In 
particular, the RG predicts the large enhancement as a function of 
$\l$ seen in the data and also predicts the dip observed in the graph 
of $\k_e$ versus $\k_0$ for sufficiently large $\l$. 
 
Indeed, the qualitative success of the method suggests that the 
difficulty in obtaining predictions that are more accurate might lie 
with the inadequacy of the simple ansatz when applied to the case when 
$\l=0$. The effect of helicity is nevertheless well captured in this 
approach because this effect is dominated by the renormalization of 
$\gb(\L)$. 
 
A technical point in the numerical integration is that $\k_0=0 
\Rightarrow \k(\infty) = 0$, and the evolution equations are 
ill-defined in the limit $\Lambda \to \infty$. This problem is easily 
rectified by making $\k_0$ very small but non-zero. The integration 
procedure is then well-defined and the results are insensitive to the 
exact value of $\k_0$ in this case. 
 
We therefore believe that although the renormalization procedure is 
not quantitatively accurate (as should be expected as it does not give 
very accurate results in the absence of helicity), it successfully 
incorporates the underlying mechanism for the enhancement of the 
diffusivity by helicity at small bare molecular diffusivity. 
 
\newsection{\label{CONC} \bf Discussion and Conclusions} 
 
In this paper we have studied the problem of turbulent advection of a 
scalar field by an incompressible flow with helicity $\l,~0 \le \l \le 
1.0$, and background molecular diffusivity, $\k_0$. We have performed 
computer simulations of the advection for flows with properties 
described in eqns. (\ref{ACF}) to (\ref{NORMS}), and compared the 
long-range effective parameters describing the time evolution of the 
scalar field with various schemes of calculation. In particular, we 
have concentrated on how the effective diffusivity, $\k_e$, depends on 
$\k_0$ and $\l$. In earlier work we found an strong anomalous 
enhancement of $\k_e$ as a function of $\l$ for $\k_0=0.0$ 
\cite{DruDuaHor:1} which was unexplained theoretically, and this is 
the motivation for the present study. In that earlier work the 
turbulent velocity field was time dependent whereas here it is 
not. This allows for easier calculation whilst still reproducing the 
effect.

The important region for discussion can be seen from the simulation 
data, figures \ref{kevk000} -- \ref{kevl02}, to be $\k_0 < 0.2$; for 
larger $\k_0$ the molecular diffusivity begins to dominate and not 
only is the effect of helicity suppressed but also the many schemes of 
calculation give good approximations for $\k_e$. For $\l=0.0$ we find 
that the two-loop self-consistent calculation of $\k_e$ reproduces the 
simulation data for all $\k_0$ very closely indeed, as is seen in 
figure \ref{kevk000} and described in section \ref{SCM}. The other 
schemes also plotted are much less accurate in the region of 
interest. Ordinary perturbation theory is not convergent in this 
region and will be discussed no further. The reason why the 
Hartree-Fock and (RG) methods are less accurate is that the vertex 
function $V(\bkpr,\bk)$, eqn. (\ref{GEN_VERTEX}), is not renormalized 
for low wavenumber which means that the associated coupling $\a$ is 
not renormalized. The Hartree-Fock method at $\l=0$ sums rainbow 
diagrams but does not include any diagrams corresponding to a vertex 
correction, unlike the self-consistent theory. In the self-consistent 
theory the one-loop prediction for $\k_0=0$ is $\k_e=1/3$ and the 
two-loop terms modify this by $\de\k_e \sim 0.04$, of which the 
two-loop cross diagram in figure \ref{KE_SC} contributes only 10\%, or 
$\de\k_e \sim 0.004$. In omitting similar terms to this latter one the 
Hartree-Fock approximation should therefore not be expected to be too 
discrepant and this is seen to be the case. The RG calculation gives a 
form which must yield the simple one-loop perturbation theory 
expression at large $\k_0$ but allows continuation to $\k_0=0$; this 
is given in eqn. (\ref{RG_L=0}). In the case of gradient flows the RG 
approach is remarkably successful \cite{DeaDruHor:1} and this is 
attributed to the fact that in that case the primitive vertex is 
renormalized at low wavenumber. 
 
The reason for examining schemes alternative to self-consistent 
methods is that for $\l > 0$ agreement between simulation data and 
theory is poor and it is necessary to investigate different approaches 
in order to test different hypotheses for a simple description of the 
observed anomalous effect. 
 
For $\k_0 > 0.2$ all schemes except ordinary perturbation theory begin 
to show reasonable agreement with the data, and for $\k_0 > 0.5$ all 
schemes clearly reproduce the results. We concentrate on results for 
$\k_0 < 0.2$ and the anomalous enhancement of $\k_e$ by helicity in 
this region is seen in figure \ref{kevl00} where $\k_e$ is plotted 
against $\l$ for $\k_0=0$, and is characterized by a rapid rise in 
$\k_e$ for $\l > 0.2$~. An alternative aspect is seen in figure 
\ref{kevk010} where $\k_e$ is plotted against $\k_0$ for $\l=1.0$. The 
significant dip at $\k_0=0.2$ is due to the large effect of helicity 
on $\k_e$ smaller $\k_0$ compared with the much reduced effect at 
$\k_0 \sim 0.2$. It has proved very difficult to convincingly explain 
these features. However, we have been able to suggest mechanisms which 
show the presence of helicity in the flow can produce a large change 
in $\k_e$ from the non-helical value, and even though these have not 
yet yielded quantitative predictions they do point towards a 
reasonable explanation. 
 
The basic idea is to recognize that the effective equation governing 
the advection should contain terms not present in the original 
equation. The terms can be thought of as being induced in the 
low-wavenumber effective theory by integrating out higher 
wavenumbers. This may also be viewed as the renormalization of the 
related vertex functions of the theory corresponding to a selective 
resummation of diagrams. In our approach we have assumed that a 
low-wavenumber approximation will be valid and so such terms will 
contain the minimum number of derivatives. These ideas can be 
implemented in different ways and we tried self-consistent, 
Hartree-Fock and renormalization group approaches. The self-consistent 
and Hartree-Fock methods are based on the effective evolution equation 
(\ref{RG_DIFFEQ}) which corresponds to a low-wavenumber enhancement of 
the flow velocity $\bu_R = \bu+\gb_R\om$, where $\om$ is the 
vorticity. We performed a two-loop calculation self-consistent in both 
$\k_e$ and $\gb_R$ as described in eqn. (\ref{SC2}) and figure 
\ref{GEN_SC}. The results show that a strong enhancement in $\k_e$ is 
predicted but that the magnitude for $\l=1.0,\k_0=0.0$ is too small 
and the form of the dependence of $\k_e$ on $\l$ disagrees with the 
data. This is particularly true at small $\l$ for $\k_0=0.0$ where 
from figure \ref{kevl00} we see that $\k_e$ is only weakly dependent 
on $\l,~\l < 0.3$ whereas we predict $\k_e \sim a+b\l^p$ for 
fractional $p$: the one-loop result is $p=2/3$. 
 
The Hartree-Fock method, shown diagramatically in figure \ref{HF}, 
computes the complete propagator in terms of $\Om(k^2)$ 
(eqn. (\ref{OMEGA})) as a sum of the rainbow diagrams generated from 
the effective equation (\ref{RG_DIFFEQ}) with $\gb_R$ given by the 
one-loop result $\gb_R = -B_1(\gb_R)/\k_e^2$ where $B_1(\gb_R)$ is 
given in eqn. (\ref{B1C1}) and $\k_e = \k_0+\Om(0)$. A rise in $\k_e$ 
with $\l$ is predicted but not one big enough to agree with the 
data. However, the behaviour at small $\l$ is a much slower rise which 
is more in keeping with the data than the self-consistent prediction 
in this region. 
 
The renormalization group method is a different approach in that it 
considers a running diffusivity $\k(\L)$ and velocity field 
$\bu(\bx,\L)$ which satisfy $\k(\infty) = \k_0,~\k(0) = 
\k_e,~\bu(\bx,\L) = \bu(\bx)$. There are three coupled RG flow 
equations, (\ref{eq:rgf}), for $\k(\L)$ and the two running spectral 
functions $\Phi(q,\L),~\Psi(q,\L)$ which correspond to the definitions 
in eqn. (\ref{F_RG}).  The important feature of the RG flow equations 
is that $\k(\L)$ appears in the denominators. The numerators are 
suppressed at large $\L$ by the spectral functions and so the major 
contribution is from intermediate values of $\L$: $\L \sim k_0$. This 
contribution is strongly enhanced for small $\k_0$ and results in the 
prediction of the dip structure observed in the data but not predicted 
by the other methods. 
 
>From our investigation we must expect that a proper explanation of the 
observed effects will require the correct effective equation and the 
consequent generation of new vertices but that unlike the $\l=0$ case 
the low-wavenumber approximation will be insufficient since although 
an enhancement is predicted for $\l \ne 0$ the rapid rise is not 
reproduced and no dip is observed. The RG method suggests that the 
main contribution is from wavenumbers $k \sim k_0$ supporting this 
latter conclusion. A successful approach should therefore include more 
terms in the effective flow equation in combination with an RG 
approach. The challenge is to obtain accurate results for all $\l$ 
including $\l=0$ by such a technique. Work in this direction is 
currently underway.

\bibliography{ref_diffusion} \bibliographystyle{unsrt}

\end{document}

%% file: head.tex
\def\ben{\begin{equation}}
\def\ba{\begin{array}}
\def\bea{\begin{eqnarray}}

\def\een{\end{equation}}
\def\eea{\end{eqnarray}}
\def\ea{\end{array}}
\def\btab{\begin{table}}
\def\btabu{\begin{tabular}}
\def\etab{\end{table}}
\def\etabu{\end{tabular}}
\def\bit{\begin{itemize}}
\def\eit{\end{itemize}}

\def\la{\langle}
\def\ra{\rangle}
\def\pd{\partial}
\def\d{\hbox{d}}
\def\a{\alpha}
\def\om{\omega}
\def\Om{\Omega}
\def\de{\delta}

\def\bom{\mbox{\boldmath $\omega$}}
\def\bxi{\mbox{\boldmath $\xi$}}
\def\bchi{\mbox{\boldmath $\chi$}}
\def\bet{\mbox{\boldmath $\eta$}}
\def\beps{\mbox{\boldmath $\epsilon$}}
\def\gb{\beta}
\def\D{\Delta}
\def\e{\epsilon}

\def\k{\kappa}
\def\L{\Lambda}

\def\l{\lambda}

\def\O{\Omega}

\def\S{\Sigma}

\def\Th{\Theta}
\def\Gt{\tilde{G}}
\def\ut{\tilde{\bu}}

\def\U{{\cal U}}
\def\half{{\textstyle{1 \over 2}}}

\def\b1{{\bf 1}}

\def\bx{{\bf x}}
\def\bxp{{\bf x^\prime}}
\def\bk{{\mbox{\boldmath k}}}
\def\bq{{\mbox{\boldmath q}}}
\def\bp{{\mbox{\boldmath p}}}
\def\bu{{\mbox{\boldmath u}}}
\def\bV{{\mbox{\boldmath V}}}
\def\bW{{\mbox{\boldmath W}}}
\def\bU{{\mbox{\boldmath U}}}
\def\bkpr{{{\bk}^\prime}}
\def\bkhat{{\hat{\bk}}}

\def\dk{{\hbox{d}\bk \over (2\pi)^3}}
\def\dq{{\hbox{d}\bq \over (2\pi)^3}}

\def\dk{{d^3k \over (2\pi)^3}}
\def\cos{\hbox{cos}\:}
\def\sin{\hbox{sin}\:}

\def\c{\cite}
\def\nn{\nonumber}
\def\bb{\left(}
\def\bbs{\left[}
\def\eb{\right)}
\def\ebs{\right]}

\newcommand{\name}{\arabic{section}}
\newcommand{\newsection}[1]{\section{#1}\renewcommand{\theequation}
                              {\name.\arabic{equation}}
                            \setcounter{equation}{0}}
\newcommand{\newsubsection}[1]{\subsection{#1}\renewcommand{\theequation}
                               {\name.\arabic{subsection}.\arabic{equation}}
                               \setcounter{equation}{0}}

\def\U{{\bf U}}

\def\u{\tilde{u}}

\def\circo{\begin{picture}(5,5)\put(3,3){\circle{5.0}}\end{picture}}

\usepackage{epsf,latexsym}

%% file: hel_rev.bbl
\begin{thebibliography}{1}

\bibitem{Kra:1}
R.H. Kraichnan.
\newblock {\em Phys. Fluids}, 13:22, 1970.

\bibitem{Kra:2}
R.H. Kraichnan.
\newblock {\em J. Fluid Mech.}, 77:753, 1976.

\bibitem{Kra:3}
R.H. Kraichnan.
\newblock {\em J. Fluid Mech.}, 81:385, 1977.

\bibitem{DruDuaHor:1}
I.T. Drummond, S~Duane, and R.R. Horgan.
\newblock {\em Nucl. Phys.}, B220:119, 1983.

\bibitem{DruHor:1}
I.T. Drummond and R.R. Horgan.
\newblock {\em Phys. Lett.}, B321:246--253, 1994.

\bibitem{DeaDruHor:1}
D.S. Dean, I.T. Drummond, and R.R. Horgan.
\newblock {\em J. Phys:A: Math Gen}, 27:5135--5144, 1994.

\bibitem{DeaDruHor:2}
D.S. Dean, I.T. Drummond, and R.R. Horgan.
\newblock {\em J. Phys:A: Math Gen}, 28:1235--1242, 1995.

\bibitem{PhyCur:1}
R.~Phythian and W.D. Curtis.
\newblock {\em J. Fluid Mech.}, 89:241, 1978.

\bibitem{DruDuaHor:2}
I.T. Drummond, S~Duane, and R.R. Horgan.
\newblock {\em J. Fluid Mech.}, 138:75--91, 1984.

\end{thebibliography}
